\documentclass[useAMS,usenatbib]{mn2e}
\usepackage{graphicx}    
\usepackage{dcolumn}     
\usepackage{bm}          
\usepackage{amssymb,amsmath}  
\usepackage{color}
\usepackage{fixltx2e}
\usepackage{hyperref}

\def\citejap#1{\citeauthor{#1}\ \citeyear{#1}}
\def\mpcoh{\,h^{-1}{\rm Mpc}}

\renewcommand{\theenumi}{(\arabic{enumi})}

\title[Halo abundances within the cosmic web]
      {Halo abundances within the cosmic web}
\author[D. Alonso et al.]
{D. Alonso$^1$\thanks{david.alonso@astro.ox.ac.uk},
 E. Eardley$^2$\thanks{ee@roe.ac.uk},
 J. A. Peacock$^2$\thanks{jap@roe.ac.uk}\\
 $^1$Astrophysics, University of Oxford, DWB, Keble Road, Oxford OX1 3RH, UK\\
 $^2$Institute for Astronomy, University of Edinburgh, Royal Observatory,
     Blackford Hill, Edinburgh EH9 3HJ, UK\\
}
\begin{document}
  \date{\today}
  \pagerange{1--13} \pubyear{2014}
  \maketitle

  \begin{abstract}
    We investigate the dependence of the mass function of dark-matter haloes on their
    environment within the cosmic web of large-scale structure. A dependence of the halo
    mass function on large-scale mean density is a standard element of cosmological
    theory, allowing mass-dependent biasing to be understood via the peak-background
    split. On the assumption of a Gaussian density field, this analysis can be
    extended to ask how the mass function depends on the geometrical environment:
    clusters, filaments, sheets and voids, as classified via the tidal tensor (the
    Hessian matrix of the gravitational potential). In linear theory, the problem can
    be solved exactly, and the result is attractively simple:
    the conditional mass function has no explicit dependence on the local tidal field, and is a
    function only of the local density on the filtering scale used to define the
    tidal tensor. There is nevertheless a strong {\it implicit\/}
    predicted dependence on geometrical environment, because the local density couples
    statistically to the derivatives of the potential. We compute the predictions of this
    model and study the limits of their validity by comparing them to results deduced
    empirically from $N$-body simulations. We have verified that, to a good approximation,
    the abundance of haloes in different environments depends only on their densities, and not
    on their tidal structure. In this sense we find relative differences between halo
    abundances in different environments with the same density which are smaller than $\sim13\%$.
    Furthermore, for sufficiently large filtering scales, the agreement with the theoretical
    prediction is good, although there are important deviations from the Gaussian prediction at
    small, non-linear scales. We discuss how to obtain improved predictions in this regime, using
    the `effective-universe' approach.
  \end{abstract}

 
  \section{Introduction}\label{sec:intro}
    Environmental modifications of the properties of galaxies and galaxy
    systems is one of the central issues in the study of the formation of
    galaxies and large-scale structure (e.g. \citejap{1965ARA&A...3....1A},
    \citejap{1996MNRAS.282..347M}, \citejap{2000MNRAS.318.1144P},
    \citejap{2002ApJ...575..587B}, \citejap{2003ApJ...593....1B}). The different observed degrees
    of clustering of galaxies of different types can be traced to this root
    cause, and the variation of galaxy bias with scale and with redshift
    can be understood quantitatively in this way.  This standard piece of
    cosmological theory has two distinct elements, which are the galaxy
    contents of different dark-matter haloes, and how the properties of
    the halo population itself can vary with location within the overall
    density field. In the majority of published work, `galaxy environment'
    refers purely to the former aspect, and is a shorthand for properties
    such as the mass of the halo hosting a given set of galaxies (e.g.
    \citejap{muldrew2012}). In this
    paper, we shall be concerned with the larger-scale question -- in
    effect defining environment via the density on scales of a few Mpc,
    rather than $<100$~kpc. Independent of the extent to which different haloes may
    generate different galaxy contents, a detailed understanding of the
    large-scale modulation of halo properties is a necessary preliminary. It has the virtue that
    it can be studied theoretically using $N$-body simulations of dark
    matter only.  Furthermore, observational analogues of haloes can be
    obtained directly via catalogues of galaxy groups, so in principle
    there is much that can be learned about this aspect of cosmological
    structure without having to address the full complexity of galaxy
    formation.

    There has been increasing recent interest in focusing the study of
    environmental modifications on what might be termed
    `geometrical environment' -- i.e. the location of a given
    galaxy or halo within the cosmic web \citep{1970Ap......6..320D,
    2007MNRAS.375..489H,2009MNRAS.396.1815F,2012MNRAS.425.2049H,
    2014arXiv1404.2206F,2015MNRAS.446.1458M,Nuza2014,2014MNRAS.441.2923C}.
    This term captures the visually striking way in which the nonlinear
    cosmic density field separates into distinct structures with different
    dimensionality, and normally four cases are considered:
    3D voids and knots at either extreme of overdensity, connected
    by 2D sheets and 1D filaments. A variety of different techniques
    have been proposed for constructing this decomposition.
    Some are theoretically meaningful, but hard to apply to
    galaxy survey data, such as the use of the peculiar velocity
    shear tensor (e.g. \citejap{2013MNRAS.428.2489L,
    2015MNRAS.446.1458M}). At the opposite extreme, there are empirical
    methods that are connected very
    closely to the nature of the data in a given survey
    (e.g. \citejap{alpaslan2014}), but which are hard to relate
    to a priori theory. There is also the question of whether
    we should be trying to study the `skeleton' of the pattern
    of the cosmic web, as in \citet{sousbie2009}, or providing
    a classification analysis of space -- so that each point
    and each astronomical object can be allocated to a unique
    geometrical environment. We take the latter approach, and
    will also focus on the effect of this environment
    on the most primitive cosmic structures: the dark-matter
    haloes. The properties of galaxies within the cosmic
    web are perhaps more readily observed, but the predictions
    are more model dependent; however, a framework such as
    the halo model is widely used to predict the galaxy
    distribution once the spatial distribution of haloes
    is known, so it makes sense to start with this fundamental problem.

    Accordingly, we summarise in Section 2 some standard results concerning the
    conditional mass function of haloes, and how this is modulated in regions of
    different large-scale overdensity; Section 3 introduces the classification
    of geometrical environment via the Hessian of the gravitational potential;
    Section 4 derives the impact of this definition of the cosmic web on the
    halo mass function; Section 5 contrasts these predictions with results
    from simulations and in Section 6 we present and discuss the main conclusions of this paper.

  \section{The conditional halo mass function}\label{sec:cmf}
    Most of the information about the non-linear accretion of dark matter
    haloes is encoded in the mass function $n(M)\,dM$, the comoving number
    density of haloes with mass $M\in(M,M+dM)$.
    \citet[PS hereafter]{1974ApJ...187..425P} developed a theoretical framework for calculating the
    mass function based solely on Gaussian statistics and non-linear spherical
    collapse. Their main result is that the collapsed fraction $F(>M)$
    (the fraction of the Universe collapsed into virialized structures of
    masses larger than $M$) can be written as a universal (cosmology independent)
    function of the variable
    \begin{equation}
      \nu_c\equiv\frac{\delta_c}{\sigma(M)},
    \end{equation}
    where $\delta_c\simeq1.686$ is the linear threshold for spherical collapse
    and $\sigma^2(M)$ is the variance of the density contrast filtered on a
    scale $R=(3M/4\pi\rho_0)^{1/3}$, given in linear perturbation theory by
    \begin{equation}
      \sigma^2(M)=\frac{1}{2\pi^2}\int_0^{\infty}P_k\,
      |W(kR)|^2\,k^2\,dk,
    \end{equation}
    where $W(kR)$ is the transform of a top-hat window function and $P_k$ is
    the matter power spectrum. The PS result is
    \begin{equation}\label{eq:ps}
      F(>M)=1-{\rm erf}\left(\frac{\nu_c}{\sqrt{2}}\right),
    \end{equation}
    from which the mass function can be calculated as
    \begin{equation}
      n(M)\,dM=\frac{\rho_0}{M}\left|\frac{dF(>M)}{d\ln M}\right|\,d\ln M.
    \end{equation}
    Note that it is often convenient to express this in the form of the
    multiplicity function: $M^2 n(M)/\rho_0$, which is the fraction of
    the mass of the Universe carried by haloes in unit range of $\ln M$.

    \subsection{The excursion set formalism}\label{ssec:exc_set}
      The PS result was given a more solid mathematical foundation by
      \citet{1991ApJ...379..440B} and by \citet{1993MNRAS.262..627L} through the
      so-called excursion set formalism. In this framework the density contrast
      at a given point as a function of the
      smoothing scale, $\delta(R)$, forms a random walk in $\delta-R$ space.
      The collapsed fraction can then be calculated as the fraction of all those
      random walks that, starting at $\delta(R\rightarrow\infty)\rightarrow0$,
      make a first crossing above the $\delta_c$ threshold at some $R>R(M)$. 
      In its original form, this formalism has one important caveat:
      for the walks to be completely random (i.e. with uncorrelated steps) the
      density contrast must be filtered using a top-hat window function in
      $k$-space, whereas the original PS result assumed top-hat filters in configuration
      space. There exist a few approaches in the literature to take into account
      the non-zero correlations between steps when using different window functions
      \citep{1990MNRAS.243..133P,2010ApJ...711..907M,2012MNRAS.427.3145M}.

      The power of the excursion set formalism is that it can be extended to study the merging
      history of haloes as well as the conditional distribution of haloes in environments with
      different large-scale densities \citep{1996MNRAS.282..347M}, since both problems
      can be addressed by studying random walks crossing two barriers of different heights.
      Assuming Gaussian statistics one can compute the conditional probability for the
      density contrast smoothed over $R_h$ to have a value $\delta_h$ given the value
      $\delta_e$ when smoothed on a scale $R_e$ (here we will use the subscripts $e$ for
      environmental quantities and $h$ for haloes):
      \begin{equation}\label{eq:cpdf}
        P(\delta_h|\delta_e)\,d\delta_h=\frac{d\nu_h}{\sqrt{2\pi(1-\varepsilon^2)}}
        \exp\left[-\frac{1}{2}\frac{(\nu_h-\varepsilon\,\nu_e)^2}{1-\varepsilon^2}\right],
      \end{equation}
      where 
      $\nu_e\equiv\delta_e/\sigma_{ee}$, 
      $\nu_h\equiv\delta_h/\sigma_{hh}$,
      $\epsilon\equiv\sigma_{eh}^2/(\sigma_{ee}\,\sigma_{hh})$ 
      and we have defined the covariance
      \begin{equation}\label{eq:sigmavar}
        \sigma_{ab}^2\equiv\frac{1}{2\pi^2}\int_0^{\infty}P_k\, 
        W_a(kR_a)\,W_b^*(kR_b)\,k^2dk.
      \end{equation}
      Notice that different window functions may have been used for
      $\delta_h$ and $\delta_e$. Equation (\ref{eq:cpdf}) is in
      fact the same result found in the unconditional case with a
      rescaling of the variable
      \begin{equation}
        \nu_h\rightarrow\frac{\nu_h-\varepsilon\,\nu_e}{\sqrt{1-\varepsilon^2}},
      \end{equation}
      and therefore the same rescaling applies to Equation (\ref{eq:ps}) to obtain
      the conditional collapsed fraction
      \begin{equation}\label{eq:cmf_es}
        F(>M|\delta_e)=1-{\rm erf}\left(\frac{\nu_c-\varepsilon\,\nu_e}
                                   {\sqrt{2\,(1-\varepsilon^2)}}\right).
      \end{equation}
      In both this and the unconditional problem, the random-walk approach
      solves the `cloud-in-cloud' issue and yields a collapse fraction that is
      simply twice the area under the tail of the one-point density
      distribution; again, this simple result does not hold in the 
      case of correlated trajectories. There exist other approaches in the
      literature that improve on this simple formalism to take into account
      the correlation between excursion steps as well as the peak nature of
      collapsed structures \citep{2013MNRAS.431.1503P}. We will not consider
      these here, since none of the trajectory-based approaches match `exact'
      numerical results.

      The original PS mass function in particular is well known to be a poor fit to $N$-body
      data, and more precise fitting formulae have been developed. These
      retain the main element of the PS philosophy by continuing to write the
      mass function with a universal dependence on
      the variable $\nu_c$. In general, this works extremely well, although in
      detail slight deviations from this universal scaling do exist
      \cite[e.g.][]{1999MNRAS.308..119S,2001MNRAS.321..372J,2007MNRAS.379.1067P,
      2008ApJ...688..709T,watson2013}. A more complex problem is what to do about the
      conditional mass function. Since the PS form for this uses the unconditional
      mass function with a change of variable from $\nu_c$ to $\nu_{\rm eff}
      \equiv(\nu_c-\varepsilon\,\nu_e)/\sqrt{1-\varepsilon^2}$, it is tempting to
      use the `observed' functional dependence of the mass function on $\nu_c$,
      substituting it by $\nu_{\rm eff}$. As we show explicitly in Section
      \ref{sec:sims_cmf}, this rescaling seems to be a good approximation only for
      large smoothing scales and mild environmental overdensities, and it breaks
      down for large values of $\delta_e$. This comes as no surprise, since the
      same approach for the conditional mass function as a means to study the
      merging histories of haloes has been shown to fail to match $N$-body data
      \citep{2008MNRAS.383..546C}.
      
      As with the original PS approach, the excursion set method
      provides a framework for relating the linearized Gaussian density field to the abundance
      of virialized objects that form in the fully non-linear density. For this reason,
      the environmental density entering Equation (\ref{eq:cmf_es}) must be not the value
      of the local physical density $\Delta_e$, but its linear extrapolation, which can
      be estimated using the prescription given by \citet{1994ApJ...427...51B}:
      \begin{equation}
        1+\Delta_e=\left(1-\frac{\delta_e}{\delta_c}\right)^{-\delta_c}.
      \end{equation}

    \subsection{The effective-universe approach}\label{ssec:eff_univ}
      The problem of the conditional mass function can also be approached from
      a different point of view. A well-known result of gravitational theory
      (which holds both in General Relativity and in the Newtonian limit) is
      that any spherically symmetric system behaves, at a fixed radius, in a manner
      equivalent to a spherical sub-region of
      a homogeneous universe with some effective cosmological parameters. Thus, an overdense
      spherical region embedded in a Friedmann-Robertson-Walker (FRW) universe
      will have a radius-time history identical to that of a different effective FRW model
      \citep{Weinberg:1972,Peebles:1993}.

      According to this result, and neglecting any error involved in
      applying it to non-spherical systems, we should be able to calculate
      an accurate conditional mass function for any degree of environmental
      overdensity, by exploiting the fact that the fitting formula for
      the unconditional mass function is claimed to be universal. 
      Given a set of effective cosmological parameters, we should then
      be able to calculate the mass function in the effective universe -- which
      is then equivalent to calculating the desired conditional mass function
      for environments of the same density. This approach has been discussed
      in previous works \citep{2003MNRAS.344..715G,2004ApJ...605....1G}, and in fact has been
      shown to be equivalent to the excursion set method in the limit of
      large smoothing scales \citep{2009MNRAS.394.2109M}. We will outline
      here the main steps of this method, and refer the reader to appendix
      \ref{app:eff_univ} for a more detailed description. 

      Since the small-scale perturbations inside the effective universe
      have their origin in the same primordial power spectrum as perturbations
      in the background FRW model, the variance of the linear overdensity field in the
      effective universe is just a rescaled version of the variance outside:
      \begin{equation}\label{eq:eff_var}
        \sigma_{\rm eff}(M)=D_g\,\sigma(M,R_e).
      \end{equation}
      Here $D_g$ is the ratio of the growth factor in the effective universe
      to the growth factor in the background cosmology, and $\sigma(M,R_e)$
      is the variance of the density field in the background corrected for the
      size of the environment (see Appendix \ref{app:eff_univ}, in particular
      Equation (\ref{eq:var_corrected}), for further details). Thus, according
      to the PS theory, the mass function in this effective universe has the same
      functional form as the universal mass function (Equation \ref{eq:ps}), using
      the rescaled variance in Equation (\ref{eq:eff_var}).

  \section{Defining the cosmic web}\label{sec:cweb}
    As we have seen, we can make predictions regarding the dependence of
    the mass function on the environment, when this is
    defined via the value of the density smoothed on a
    large scale -- i.e. we can ask questions about the mass
    function in giant voids, and how this compares to the result within superclusters. 
    For sharp $k$-space filtering, $\sigma_{eh}$ is identical to $\sigma_{ee}$, and
    in the limit of large $R_e$, the correlation parameter $\varepsilon$ becomes small.
    Thus the variable $\nu_{\rm eff}$ in the excursion set conditional mass function of
    Equation (\ref{eq:cmf_es}) takes the value
    \begin{equation}
      \nu_{\rm eff} = \frac{\delta_h-\delta_e}{\sigma_{hh}};
    \end{equation}
    this is the peak-background split, in which the large-scale density fluctuation is treated as
    an external offset to the density. This shift makes it easier to
    attain the $\delta_h=\delta_c$ threshold and for more haloes to collapse in high-density
    regions. But although this phenomenon is of great interest in 
    explaining the origin of biased clustering, it is not
    necessarily the best way of classifying environments if
    we actually want to observe the peak-background split in action.
    Examination of images of redshift surveys or $N$-body simulations
    shows that the majority of the mass condenses into geometrical
    features with low dimensionality and a small filling factor:
    the cosmic web. To understand biased clustering in detail, it will
    be interesting to look at how the halo population changes within
    these different environments. This means that we need to think
    about more than the mean density defined with some large
    isotropic filter; such an approach ignores e.g. the fact that
    filaments are relatively narrow features in some directions.

    A number of approaches have been taken to the problem of making a geometrical
    classification of different parts of the cosmic web
    \citep{2007MNRAS.375..489H,2009MNRAS.396.1815F,2012MNRAS.425.2049H}; we have
    chosen to follow \citet{2009MNRAS.396.1815F} and use the tidal (or deformation)
    tensor. This is defined as $T_{ij}\equiv\partial_i\partial_j\tilde{\phi}$, i.e.
    the Hessian of the Newtonian potential\footnote{In this paper all spatial
    derivatives $\partial_i$ are defined with respect to physical coordinates ${\bf r}$
    and not comoving ones ${\bf q}\equiv{\bf r}/a$. This removes the proportionality
    factor $a^2$ from the relation between $\tilde{\phi}$ and $\delta$ and makes the
    notation less cumbersome.}, and therefore the density contrast is directly given by
    Poisson's equation as its trace $\delta=T_{11}+T_{22}+T_{33}$. Note that we have normalized
    the usual definition of the Newtonian potential
    \begin{equation}
      \tilde{\phi}\equiv \phi/(4\pi G\bar{\rho}),
    \end{equation}
    to make this relation simpler.

    We will classify a given point in space as belonging to one of the different
    elements of the cosmic web according to the number of eigenvalues of
    $\hat{T}$ above a given threshold $\lambda_{\rm th}$ at that point. Thus, let
    $\lambda_1\ge\lambda_2\ge\lambda_3$ be the three eigenvalues of $\hat{T}$. 
    We define:
    \begin{itemize}
      \item {\bf Voids}: all eigenvalues below the threshold 
            ($\lambda_1<\lambda_{\rm th}$).
      \item {\bf Sheets}: only one eigenvalue above the threshold 
            ($\lambda_1>\lambda_{\rm th}$, $\lambda_2<\lambda_{\rm th}$).
      \item {\bf Filaments}: two eigenvalues above the threshold 
            ($\lambda_2>\lambda_{\rm th}$,$\lambda_3<\lambda_{\rm th}$).
      \item {\bf Knots}: all eigenvalues above the threshold 
            ($\lambda_3>\lambda_{\rm th}$).
    \end{itemize}

    In this approach, the eigenvalue threshold is an arbitrary quantity. One might be
    tempted to set it to zero and classify based on the sign of the
    eigenvalues, but this is not very attractive from the point of view
    of gravitational collapse. We have seen that the density contrast is
    the sum of the eigenvalues, and we know that nonlinear collapse
    involves a linear contrast of order unity. It is therefore natural
    to consider $\lambda_{\rm th}$ of order unity; we refine this choice below.

    A different prescription, the so-called V-web approach \citep{2012MNRAS.425.2049H}, uses
    instead the eigenvalues of the velocity shear tensor
    $\Sigma_{ij}\equiv-(\partial_iv_j+\partial_jv_i)/2\,H$. Although
    this prescription allows for a finer discrimination on very small
    scales, $\Sigma_{ij}$ converges to $f_g\,T_{ij}$ 
    in linear theory\footnote{Note also that the normalised potential $\tilde{\phi}$
    is also linearly related to the potential of the displacement vector
    ${\bf \Psi}\equiv\nabla\phi_d$ via $\tilde{\phi}= a^2\phi_d$.} (where
    $f_g\equiv d \ln\delta/d \ln a$), and therefore
    similar results should hold. Note that in practice we would need to define
    the web using a redshift survey. If we work from the galaxy density
    and use a quasi-potential obeying $\nabla^2\tilde{\phi} = -1 +
    \rho_g/\langle\rho_g\rangle$, the Hessian approach can still be used
    -- but the velocity shear tensor is not observable.

    It is useful to check that the above description makes intuitive
    sense, via a couple of analytic examples. First let us consider a 
    spherical model with $\tilde{\phi}=-a/(r^2+b^2)^{1/2}.$
    For this, $T_{ij}=a(r^2+b^2)^{-5/2}[(r^2+b^2)\delta_{ij}-3x_ix_j]$,
    and we can assess the eigenvalues by looking at the diagonal form
    in the frame $(x,y,z)=(r,0,0)$. For $r<b/\sqrt{2}$, there are three
    positive eigenvalues that coincide at $r=0$, so there is a knot
    there if $a$ is large enough (otherwise it is a void everywhere).
    Since the first diagonal entry is smaller, this spherical knot must be
    surrounded by a shell where the field is classified as a filament, with
    classification changing to a void at large enough $r$ (even though the
    density perturbation is positive everywhere). The prediction of a
    filament region is a little surprising, but sensible: away from the
    centre of the perturbation, tidal forces act on an extended object by
    `squeezing' it in the two transverse directions and stretching it in
    the radial one. More intuitive results arise from
    lower-dimensional versions of the same model: 
    $\tilde{\phi}=-a/(x^^2+x^2+b^2)^{1/2}$ makes a transition from a
    filament to a void, whereas $\tilde{\phi}=-a/(x^2+b^2)^{1/2}$
    makes a transition from a sheet to a void. In general terms, tidal
    forces act on an extended object through the second derivatives of $\tilde{\phi}$:
    \begin{equation}
      \ddot{\delta x}_i=-T_{ij}\,\delta x_j,
    \end{equation}
    and hence this prescription classifies different regions according to
    the number of directions in which tidal forces tend to contract or
    stretch an extended object.

  \section{Gaussian statistics in the cosmic web}\label{sec:stats}
    It is interesting to study the statistics of the cosmic web
    in the limit that $\tilde{\phi}$ is a Gaussian random field. We
    will also assume that the potential is smoothed on a scale $R_e$
    with some window function $W_e(kR_e)$. In this case the probability density
    function of the eigenvalues $\lambda_i$ is given by:
    \begin{align}
      dP(\lambda_i) \equiv &\, p(\rho,\theta,\nu)\,d\rho\,d\theta\,d\nu\\\nonumber
               \equiv &\, 225\sqrt{\frac{5}{2\pi}}\rho\,(\rho^2-\theta^2)\,
      \exp\left[-\frac{1}{2}(15\rho^2+5\theta^2)\right]\\\label{eq:pdf_lambdas}
      &\times \frac{e^{-\nu^2/2}}{\sqrt{2\pi}}\,d\rho\,d\theta\,d\nu,
    \end{align}
    where we have defined the variables
    $\nu\equiv(\lambda_1+\lambda_2+\lambda_3)/\sigma_{ee}\equiv\delta_e/\sigma_{ee}$,
    $\rho\equiv(\lambda_1-\lambda_3)/2\sigma_{ee}$ and
    $\theta\equiv(\lambda_1-2\lambda_2+\lambda_3)/2\sigma_{ee}$. The derivation of 
    this result is outlined in Appendix \ref{app:cweb} and has been widely
    used in the literature \citep{1970Ap......6..320D,1986ApJ...304...15B,
    1996ApJS..103....1B,2013MNRAS.430.1486R}. Note that these results are usually
    quoted in terms of the ellipticity $e$ and prolateness $p$ which are related
    to our variables by $e\equiv\rho/\nu$, $p\equiv\theta/\nu$.

    The restriction $\lambda_1\ge\lambda_2\ge\lambda_3$ implies $\rho\in[0,\infty)$,
    $\theta\in[-\rho,\rho]$, and by selecting a given type of environment we further
    constrain the dynamical range of $\nu$. Let $\lambda_{\rm th}$ be the eigenvalue
    threshold used for our classification, and let $\alpha$ denote the number of
    eigenvalues above this threshold for each case. Then the integration limits for
    $\nu$ are $f_1^{\alpha}(\rho,\theta)<\nu-\nu_{\rm th}<f_2^{\alpha}(\rho,\theta)$,
    with $\nu_{\rm th}\equiv3\lambda_{\rm th}/\sigma_{ee}$ and
    \begin{align}
      f_1^{\alpha}(\rho,\theta)=\left\{\begin{array}{ll}
                                   -\infty,       &\alpha=0\,\,\text{(voids)}\\
                                   -3\rho-\theta, &\alpha=1\,\,\text{(sheets)}\\
                                   2\theta,       &\alpha=2\,\,\text{(filaments)}\\
                                   3\rho-\theta,  &\alpha=3\,\,\text{(knots)}\\
                                  \end{array}\right.,\\
      f_2^{\alpha}(\rho,\theta)=\left\{\begin{array}{ll}
                                   -3\rho-\theta, &\alpha=0\,\,\text{(voids)}\\
                                   2\theta,       &\alpha=1\,\,\text{(sheets)}\\
                                   3\rho-\theta,  &\alpha=2\,\,\text{(filaments)}\\
                                   \infty,        &\alpha=3\,\,\text{(knots)}\\
                                  \end{array}\right..
    \end{align}

    The volume fraction in environments of type ($\alpha,\,\nu_{\rm th}$) can be
    calculated as the probability for a random point in space to belong to that type:
    \begin{equation}\label{eq:vfrac}
      F_V(\alpha,\nu_{\rm th})\equiv\int_0^{\infty}d\rho\int_{-\rho}^{\rho}d\theta
      \int_{\nu_{\rm th}+f_1^{\alpha}(\rho,\theta)}^{\nu_{\rm th}+f_2^{\alpha}(\rho,\theta)}d\nu\,
      p(\rho,\theta,\nu).
    \end{equation}
    Notice that the volume fraction is a universal function of the normalized threshold
    $\nu_{\rm th}$.

    \subsection{Dependence on the environmental density}\label{ssec:cweb_restricted}
      We aim to give a prediction for the abundance of haloes of different mass in
      different types of environment. In the excursion set approach, this abundance 
      is given by the number of first upcrossings of the collapse threshold by
      the density contrast field smoothed over a scale $R_h$, corresponding to the
      halo mass. Therefore we first need to study the joint probability for the 
      density contrast $\delta_h$ smoothed over this scale and the environmental
      tidal tensor eigenvalues, filtered on a scale $R_e$. As shown in 
      Appendix \ref{app:cweb}, this distribution is:
      \begin{equation}
        \frac{dP(\delta_h,\lambda_i)}{d\nu_h\,d\nu_e\,d\rho\,d\theta}=
        \frac{p(\rho,\theta,\nu_e)}{\sqrt{2\pi(1-\varepsilon^2)}}
        \exp\left[-\frac{(\nu_h-\varepsilon\,\nu_e)^2}{2\,(1-\varepsilon^2)}\right],
      \end{equation}
      where $\nu_e\equiv\delta_e/\sigma_{ee}$.

      As a first step, we are interested in the abundance of haloes within environments
      of type $(\alpha,\nu_{\rm th})$ with local environmental density contrast
      $\delta_e/\sigma_{ee}\in(\nu_e,\nu_e+d\nu_e)$. In this case we integrate over $\rho$
      and $\theta$ to find the joint distribution
      \begin{align}
        \nonumber
        P(\nu_h,\nu_e,\alpha,\nu_{\rm th})\,d\nu_e\,d\nu_h =  
        \frac{d\nu_e}{\sqrt{2\pi}}\,C(\nu_e)
        \,e^{-\nu_e^2/2}\\
        \times\frac{d\nu_h}{\sqrt{2\pi(1-\varepsilon^2)}}\exp\left[-
        \frac{(\nu_h-\varepsilon\,\nu_e)^2}{2\,(1-\varepsilon^2)}\right].
      \end{align}
      Here
      \begin{equation}
        C(\nu_e)\equiv225\sqrt{\frac{5}{2\pi}}\iint_{\mathcal{A}(\nu_e)}d\rho\,d\theta\,
        \rho\,(\rho^2-\theta^2)\,e^{-\frac{1}{2}\left(15\rho^2+5\theta^2\right)}
      \end{equation}
      and $\mathcal{A}(\nu_e)$ is the region in the plane $(\theta,\rho)$ defined by the
      conditions
      \begin{align}
        \nonumber
        \rho\ge0,\,\,\,&-\rho\le\theta\le\rho,\\
        f_1^{\alpha}(\rho,\theta)\le\nu_e&-\nu_{\rm th},\le f_2^{\alpha}(\rho,\theta).
      \end{align}
      On the other hand, the probability of finding such an environment is given by
      \begin{equation}\label{eq:cond_pdf}
        P(\nu_e,\alpha,\nu_{\rm th})\,d\nu_e = \frac{d\nu_e}{\sqrt{2\pi}}\,C(\nu_e)\,
        e^{-\nu_e^2/2},
      \end{equation}
      which cancels out when computing the conditional probability
      \begin{equation}
        P(\nu_h|\nu_e,\alpha,\nu_{\rm th})\,d\nu_h=P(\delta_h|\delta_e)\,d\delta_h,
      \end{equation}
      where $P(\delta_h|\delta_e)$ is given in Equation (\ref{eq:cpdf}).

      From this we can extract a general prediction of this formalism,
      which is a key result of this paper: the only environmental quantity
      that determines the abundance of haloes is the local density $\delta_e\equiv
      \lambda_1+\lambda_2+\lambda_3$. This conclusion arises from the fact that our treatment is
      based on the spherical top-hat collapse, which disregards all couplings of
      the halo orientations with other combinations of the eigenvalues of the
      tidal tensor (i.e. the only non-zero, non-diagonal element of the
      covariance matrix is between $\nu_h$ and $\nu_e$). While this
      assumption of zero coupling of gravitational collapse to tidal forces
      could be challenged in detail, we find it striking that the
      geometrical environment is not predicted to have a more direct influence
      on the properties of haloes at a given overdensity.

      In the remainder of this paper, we therefore subject our result to the
      test of confrontation with numerical simulations. Any evidence for 
      an explicit dependence on geometrical environment would be interesting
      as it would relate to the issue of halo `assembly bias'
      (e.g. \citejap{Gao2007}). This term stands for the possibility
      that halo properties have some dependence on parameters beyond
      the local overdensity; the concept may apply either to the final-state
      properties of the halo or to its merger history (which potentially
      influences the galaxy contents of the halo). This issue is
      certainly under active consideration from the point of view
      of the observational dependence of galaxy properties on tidal
      forces (e.g. \citejap{Yan2013}). It is worth noting that our analysis
      is focused on overall halo abundances and not intrinsic halo properties,
      and hence it does not directly address the problem of assembly bias.

    \subsection{The four mass functions}\label{ssec:cweb_unrestricted}
      Using the key result of the previous subsection, 
      any halo statistic conditional to a given type of environment
      can be calculated within the excursion set
      formalism as the average of that statistic conditional to a given
      background density in that environment. E.g. for the conditional collapsed
      fraction:
      \begin{align}
        \nonumber
        F(>M|\alpha,\nu_{\rm th})=&\int_0^{\infty}d\rho\int_{-\rho}^{\rho}d\theta
        \int_{\nu_{\rm th}+f_1^{\alpha}(\rho,\theta)}^{\nu_{\rm th}
                   +f_2^{\alpha}(\rho,\theta)}d\nu_e \\
        \times&\frac{p(\rho,\theta,\nu)\,F(>M|\delta_e\equiv\sigma_{ee}\nu_e)}
        {F_V(\alpha,\nu_{\rm th})},
      \end{align}
      where $F(>M|\delta_e)$ is obtained through either of the methods
      outlined in Section \ref{sec:cmf}.

  \section{Comparison with simulations}\label{sec:sims}
      \begin{figure*}
        \centering
        \includegraphics[width=0.53\textwidth]{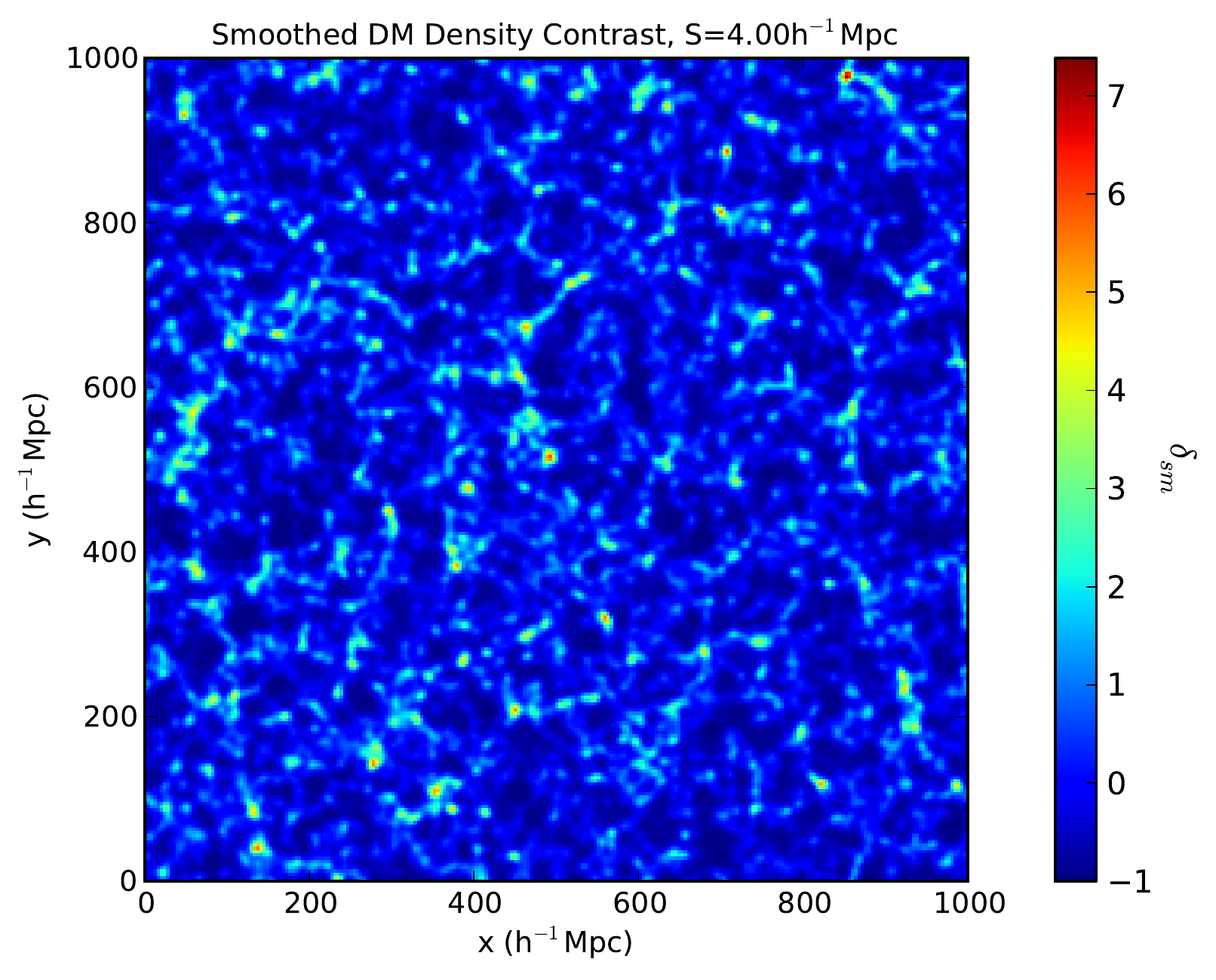}
        \includegraphics[width=0.44\textwidth]{Env_t0p40_4p00Smoothed_256grid1000}
        \caption{Density field (left) and environment classification (right) of a slice of
                 the MDR1 simulation. The colour code is voids (red); sheets (green);
                 filaments (blue); knots (yellow). A Gaussian filter with $S=4\mpcoh$ and a
                 threshold $\lambda_{\rm th}=0.4$ were used.}
        \label{fig:slices}
      \end{figure*}
    \begin{figure}
      \begin{center}
        \includegraphics[width=0.47\textwidth]
                        {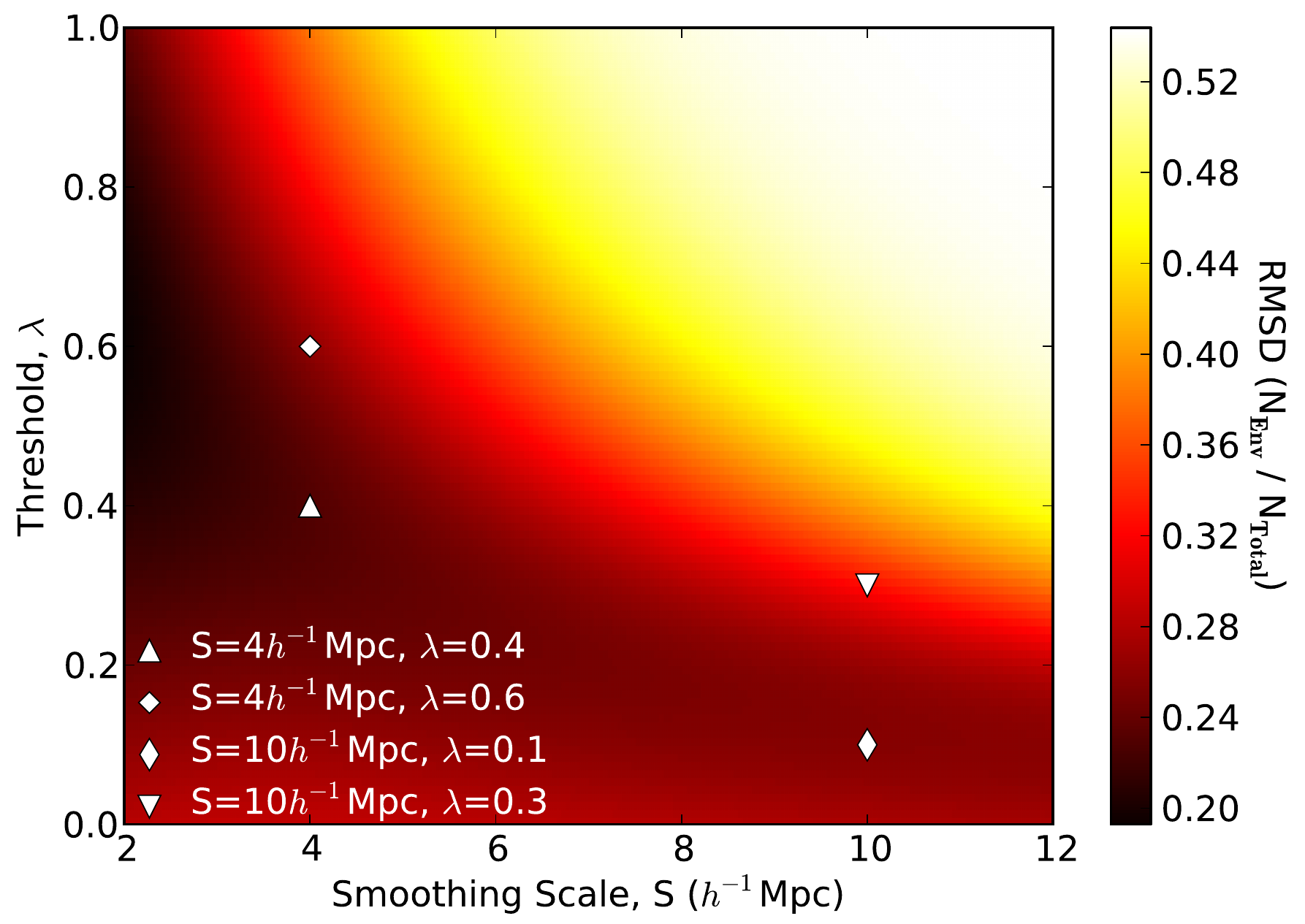}
      \end{center}
      \caption{Illustrating how the partition of the cosmic mass density between different
               geometrical environments varies with smoothing scale and threshold.
               A practically useful partition will place approximately equal quantities of
               mass in the four environments, and the colour scale shows the dispersion in
               these mass fractions. The optimum is approximately $\lambda=0.4$ for
               $S= 4\mpcoh$ and $\lambda=0.1$ for $S= 10\mpcoh$ (shown as points); but we
               also consider other thresholds, 0.6 and 0.3 respectively, to illustrate how
               our results depend on the choice of these parameters.} \label{fig:rmsd}
    \end{figure}
    \begin{figure*}
      \centering
      \includegraphics[width=0.49\textwidth]{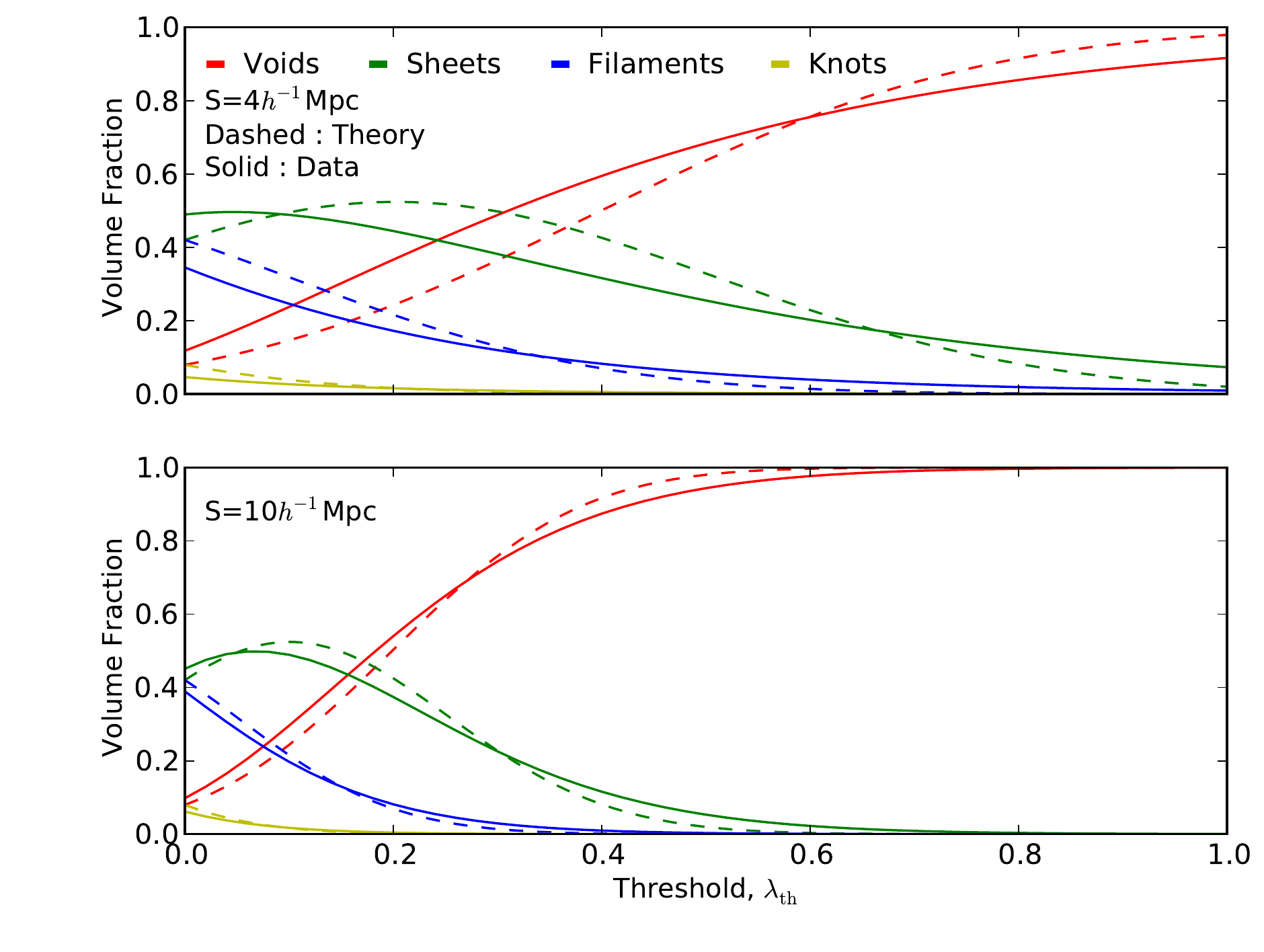}
      \includegraphics[width=0.49\textwidth]{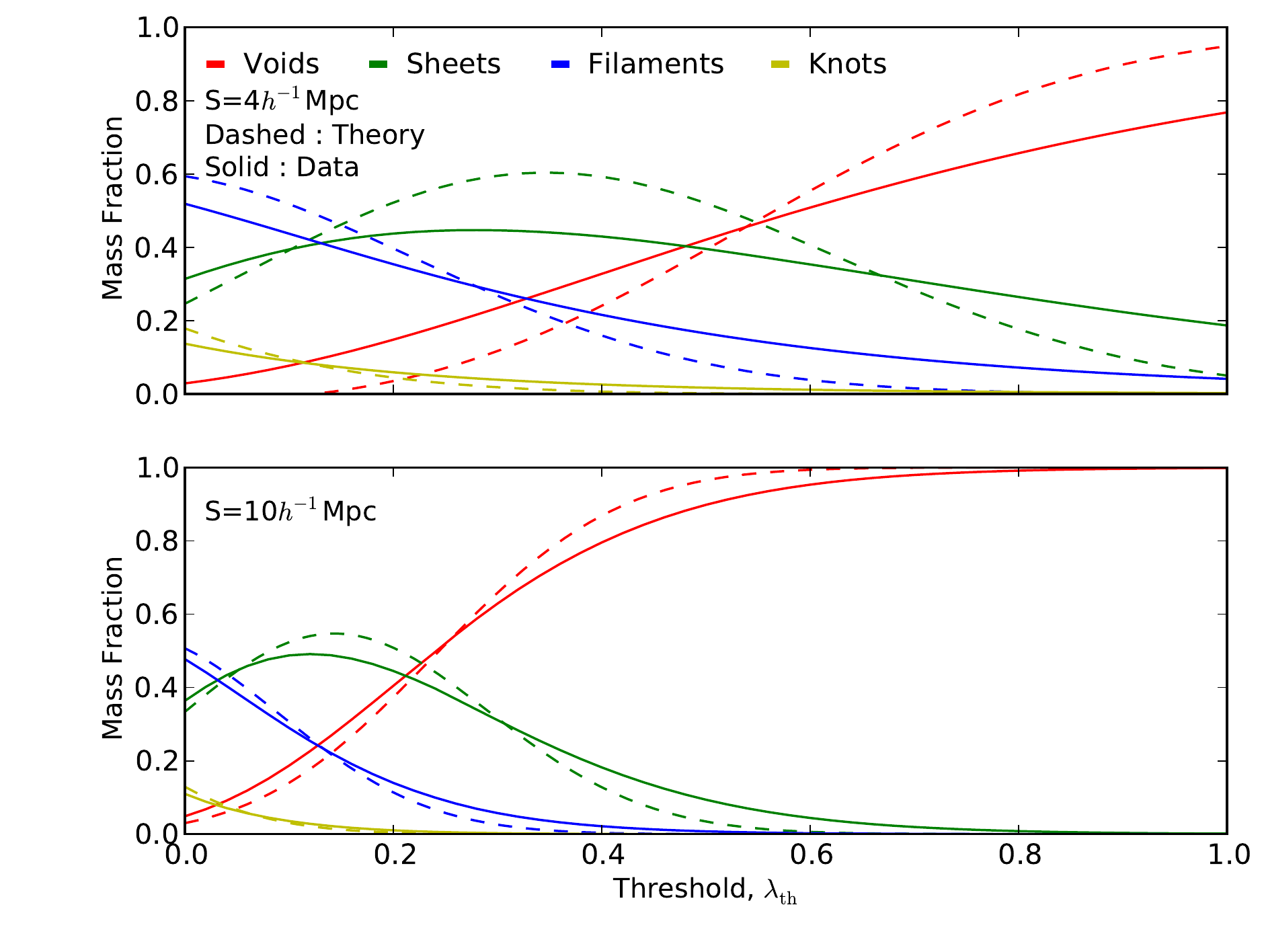}
      \caption{Volume (left) and mass (right) fractions in the four environments for
               $S=4,\,10\mpcoh$ as a function of the eigenvalue threshold, measured
               from the MDR1 simulation (dashed lines). The results are shown for
               voids (red), sheets (green), filaments (blue) and knots (yellow). The
               Gaussian-field prediction for the volume fractions are also shown as
               solid lines for comparison.}
               \label{fig:fracs}
    \end{figure*}

      \begin{figure*}
        \centering
        \includegraphics[width=0.48\textwidth]
        {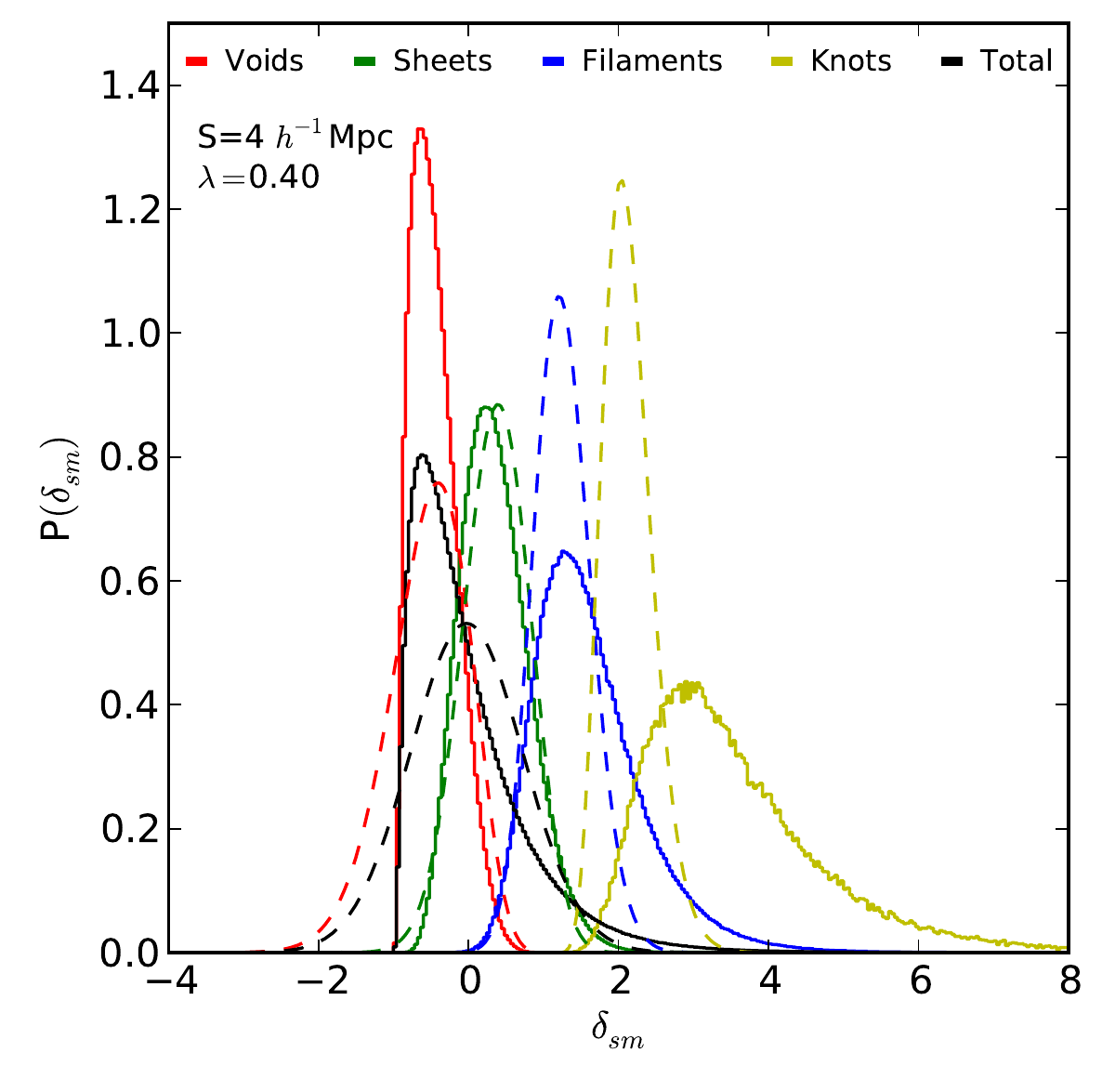}
        \includegraphics[width=0.48\textwidth]
        {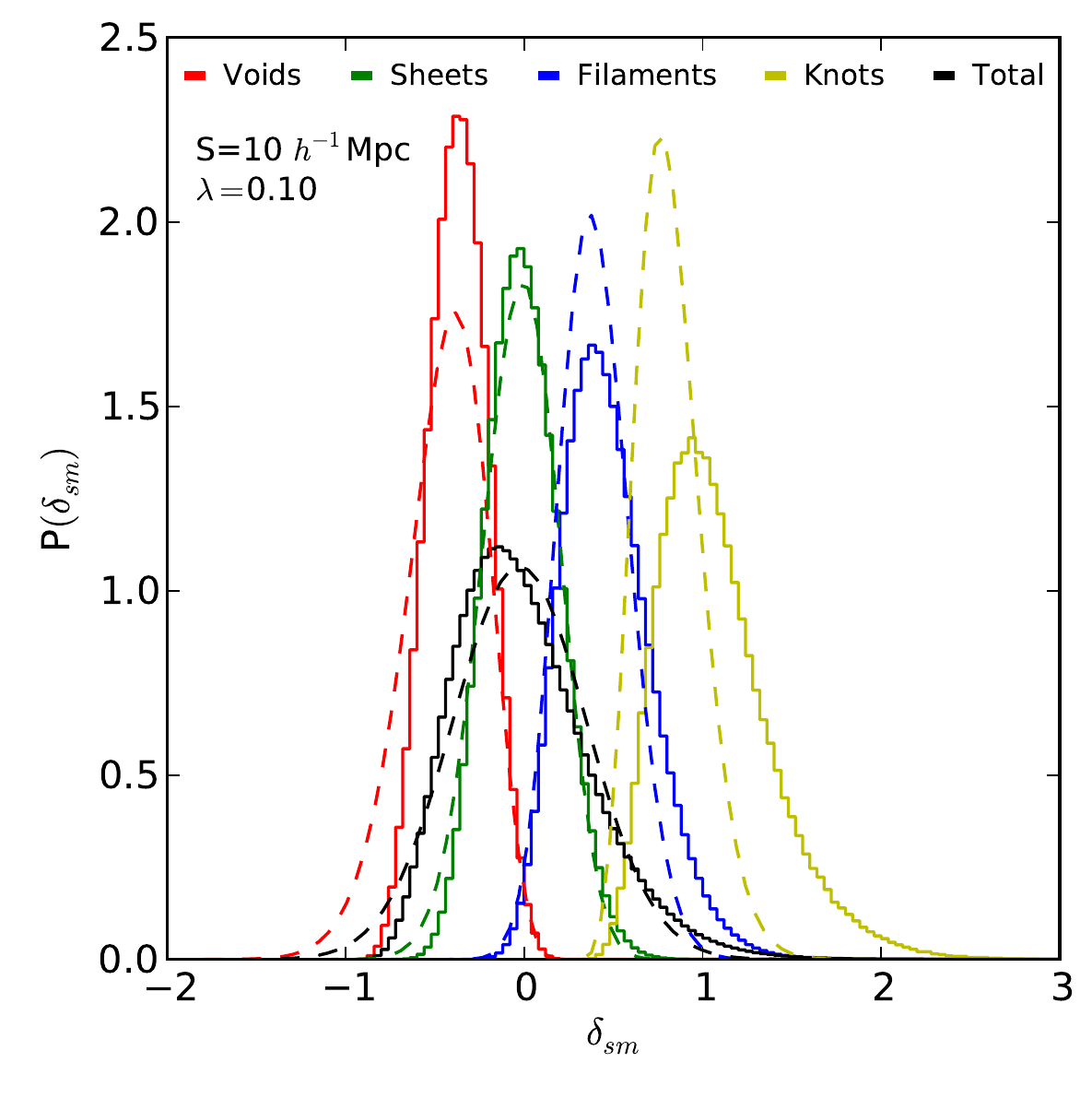}
        \includegraphics[width=0.48\textwidth]
        {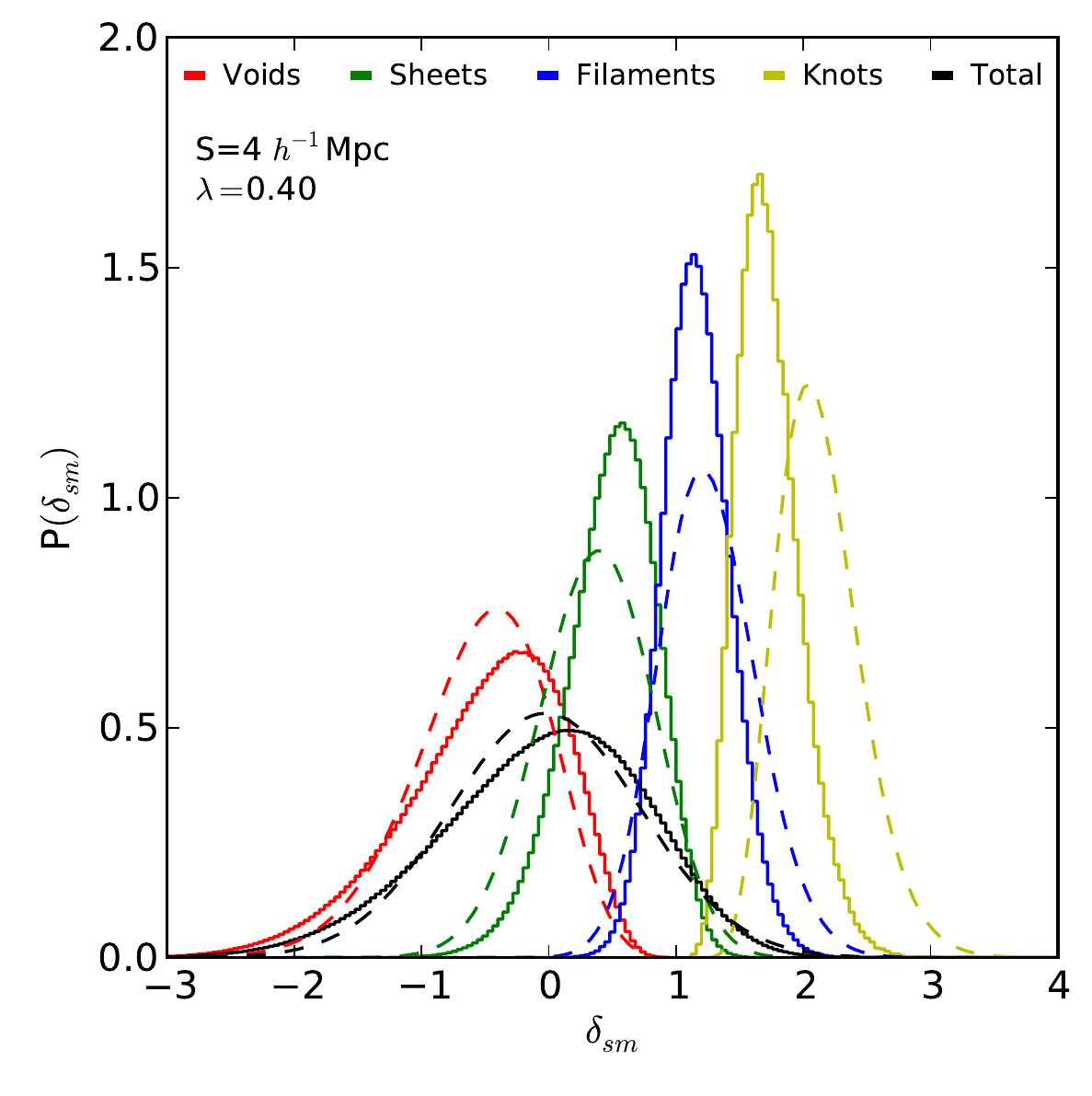}
        \includegraphics[width=0.48\textwidth]
        {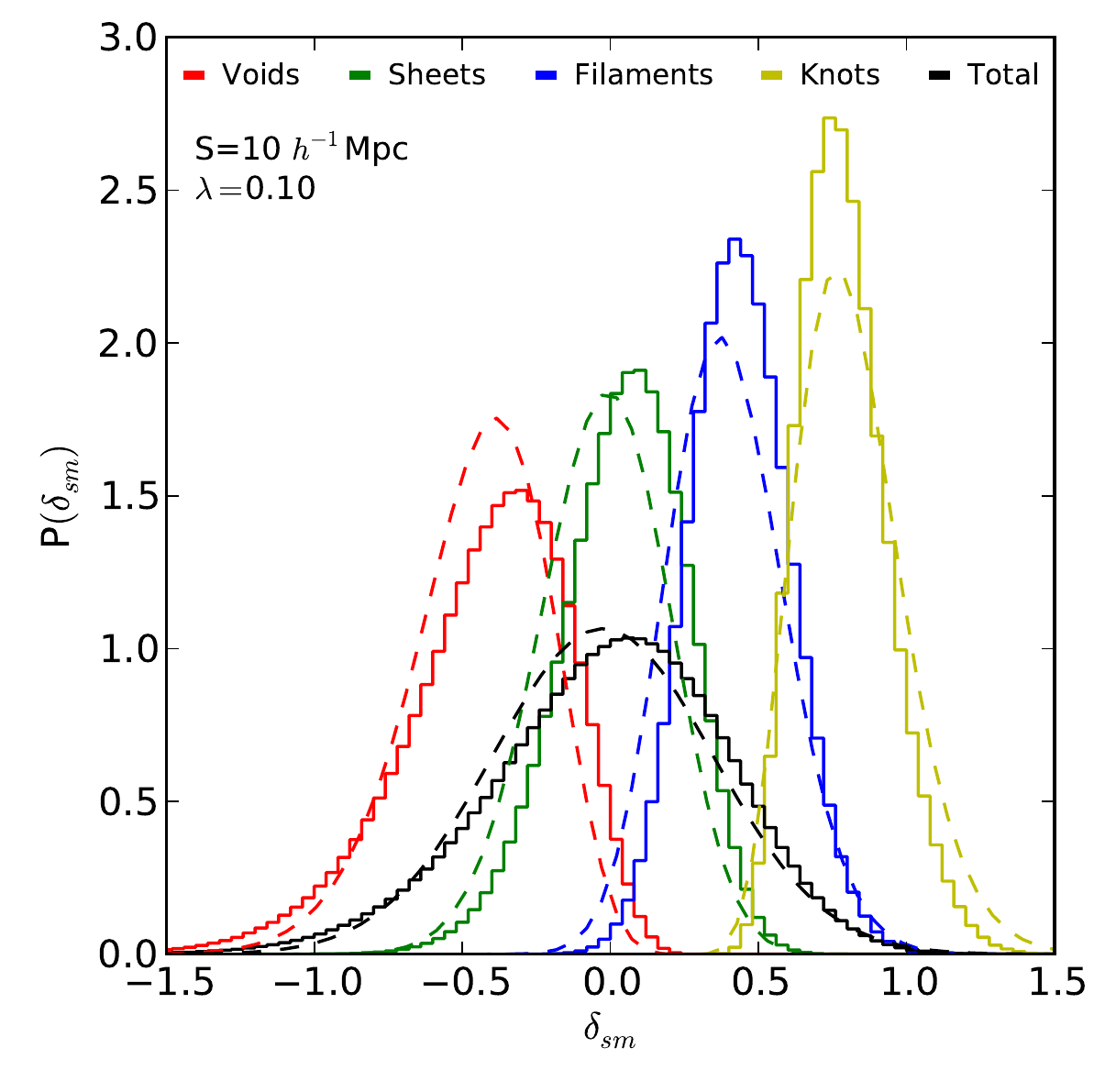}
        \caption{Overdensity distributions in each of the four environments and the overall
                 overdensity distribution for $(S,\lambda_{\rm th})=(4\mpcoh,0.4)$ (left
                 panels) and $(S,\lambda_{\rm th})=(10\mpcoh,0.1)$ (right panels). The dashed
                 lines show the Gaussian theoretical prediction, while the solid histograms show
                 the distributions extracted from the MDR1 simulation. In the top panels
                 these histograms correspond to the distribution of the real density field,
                 while the bottom panels show the distribution of the `de-lognormalized'
                 overdensity (see Equation (\ref{eq:delog})). The colour code is voids (red);
                 sheets (green); filaments (blue); knots (yellow); overall distribution (black).
                 In the same order, the density distribution for the 4 environments peaks on
                 increasing values of $\delta_{sm}$.}
        \label{fig:delta_dist}
      \end{figure*} 
    The analytical predictions explained in the previous sections have been
    compared with numerical data from the MultiDark Run 1 (MDR1) dark matter
    N-body simulation \citep{2012MNRAS.423.3018P}. MDR1 simulates a 
    $1\,(h^{-1}{\rm Gpc})^{3}$ cubic volume with a mass resolution of
    $m_{p} = 8.721\times10^{9}\,h^{-1}M_{\odot}$ in a $\Lambda$CDM cosmology
    with $(\Omega_{m}=\Omega_{dm}+\Omega_{b}, \Omega_{\Lambda}, \Omega_{b}, h, 
    n, \sigma_{8}) = (0.27, 0.73, 0.0469, 0.7, 0.95, 0.82)$. The corresponding
    halo catalogue was compiled using a friends-of-friends algorithm on the
    $z=0.1$ snapshot, yielding a minimum mass cut of $M_{\rm min}=10^{11.5}\,
    h^{-1} M_{\odot}$.

    The density field used to classify the different environments was computed
    by interpolating the dark-matter particle content on to a grid of spacing
    $a=3.9\,h^{-1}\text{Mpc}$ by cloud in cell interpolation. This field
    was then smoothed using a Gaussian filter of variable width:
    \begin{equation}
      W(k)=\exp(-S^2 k^2/2);
    \end{equation}
    we denote the width of the Gaussian by $S$, since the common symbol $\sigma$
    is used elsewhere. We choose to use Gaussian filtering for numerical stability
    in preference to spatial top-hat filtering. The results are closely
    equivalent to using a top-hat of radius $R\simeq\sqrt{5}S$. We show results for two
    different filtering scales, $S=4$ and $10\mpcoh$. The deformation tensor
    was calculated at each point in the grid by solving Poisson's equation in
    Fourier space and then transforming back to configuration space (making
    extensive use of Fast Fourier Transforms throughout the process). Fig.
    \ref{fig:slices} shows an example of the density field and environment
    classification obtained for MDR1.

    \subsection{The eigenvalue threshold}\label{ssec:eigval}
      The value of the eigenvalue threshold to be used in this type of analyses
      is somewhat arbitrary. One may choose $\lambda_{\rm th}=0$ on
      the basis that this value discriminates between purely compressing or
      stretching tidal forces. This choice, however, yields a low abundance of
      voids, compared to what one would expect from a visual examination of a
      large-scale structure map, and previous studies \cite[e.g.][]{2009MNRAS.396.1815F,
      2015MNRAS.446.1458M} have chosen a threshold in order to yield a better
      match to the visually expected volume fractions. Ultimately we would like to
      be able to extract the maximum amount of information from the abundance of
      haloes in these four environments. While it is possible to improve the halo
      statistics by resampling technicques such as the one proposed by \citet{2012arXiv1210.7871A},
      for a limited amount of data it is possible to use the value of the
      eigenvalue threshold to make sure that we obtain sufficient statistics for
      all of the different environments. In order to do so we have computed the fraction of the
      total  halo mass contained in each environment for different values of $S$ and
      $\lambda_{\rm th}$, $FM_{\alpha}(S,\lambda_{\rm th})$, and the root mean square deviation
      of these fractions $FM_{\rm rms}(S,\lambda_{\rm th})=\left[\sum_{\alpha}\,(FM_{\alpha}-
      \overline{FM})^2/4\right]^{1/2}$. We have then chosen the optimal value of
      $\lambda_{\rm th}$ for each smoothing scale $S$ as the one that minimizes
      $FM_{\rm rms}$. Besides this consideration, it is also important to make sure that the
      values of $\lambda_{\rm th}$ used yield physically sensible definitions
      for the different environments. One can then try to combine the aforementioned
      method with extra requirements, such as limiting the fraction of cells classified as
      voids that are overdense.
      
      Fig. \ref{fig:rmsd} shows the values of $FM_{\rm rms}$ for different
      smoothing scales and eigenvalue thresholds computed from the simulation
      (left panel). In view of this figure we have chosen two eigenvalue 
      thresholds for each of the two smoothing scales:
      \begin{align}
        &S=4\mpcoh\,\longrightarrow \lambda_{\rm th} = 0.4,\,0.6,\\\label{eq:envirs}
        &S=10\mpcoh\,\longrightarrow \lambda_{\rm th} = 0.1,\,0.3.
      \end{align}
      In both cases the first value was chosen by restricting the fraction of overdense void
      cells to be smaller than 10\%, while we have chosen a second slightly larger threshold
      to illustrate how our results depend on this choice. Note also that the combinations
      $(S,\lambda_{\rm th})=(10\mpcoh,0.1)$ and $(4\mpcoh,0.4)$ are similar to the values used
      in the analysis of the Galaxy And Mass Assembly (GAMMA) data by
      \citet{2014arXiv1412.2141E}.

      It is useful to verify the meaning of this choice by looking at the
      volume and mass fractions in each environment as a function of
      $\lambda_{\rm th}$. This is shown in Fig. \ref{fig:fracs}. As can be
      seen, our choice of eigenvalue threshold tries to maximize statistics of
      the four environments simultaneously. The figure also shows the Gaussian-field
      prediction for the volume and mass fractions within the formalism
      described in Section \ref{sec:stats}. The match is rather good for
      the larger value of the smoothing scale.

      \begin{figure*}
        \centering
        \includegraphics[width=0.48\textwidth]
                        {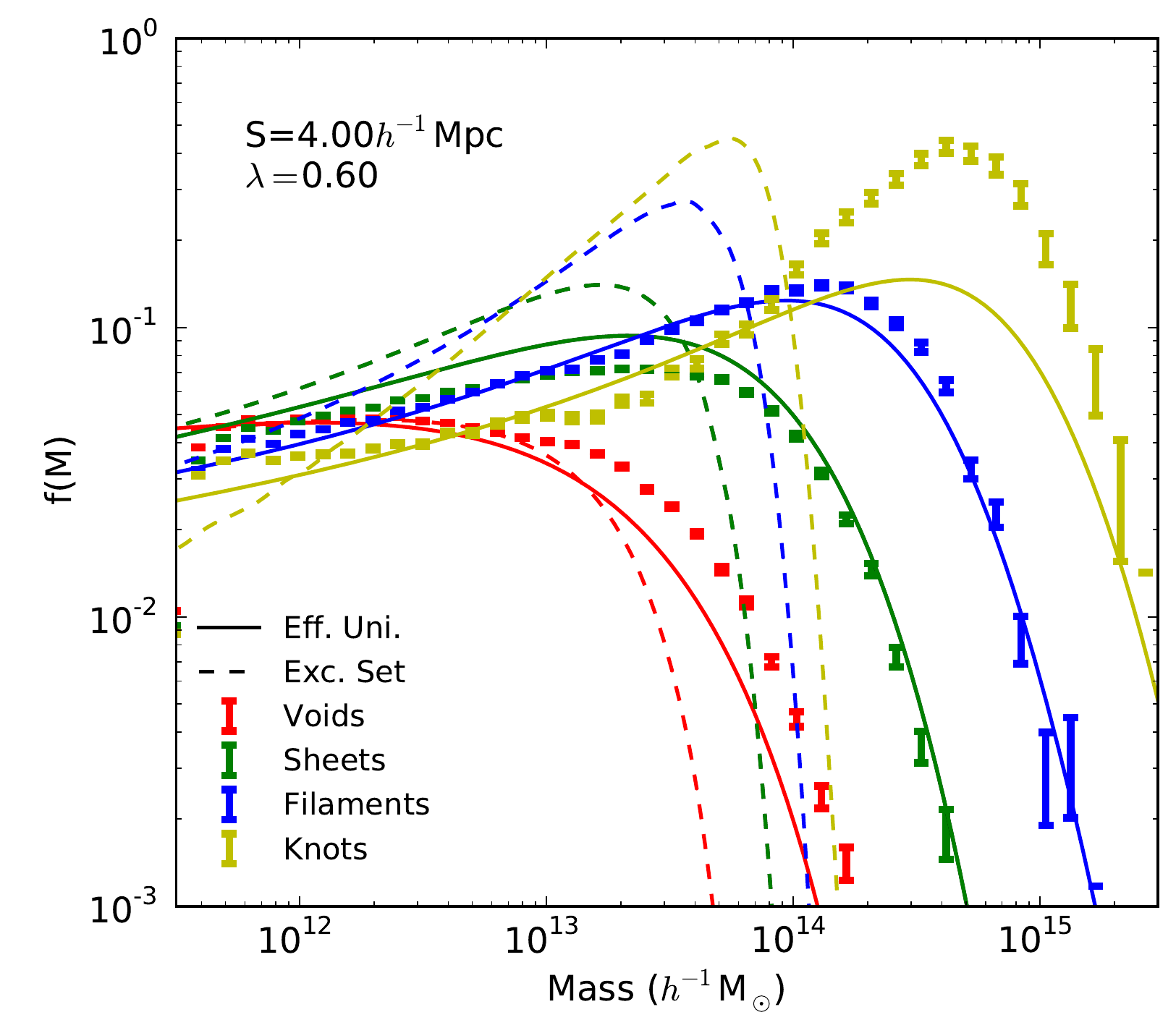}
        \includegraphics[width=0.48\textwidth]
                        {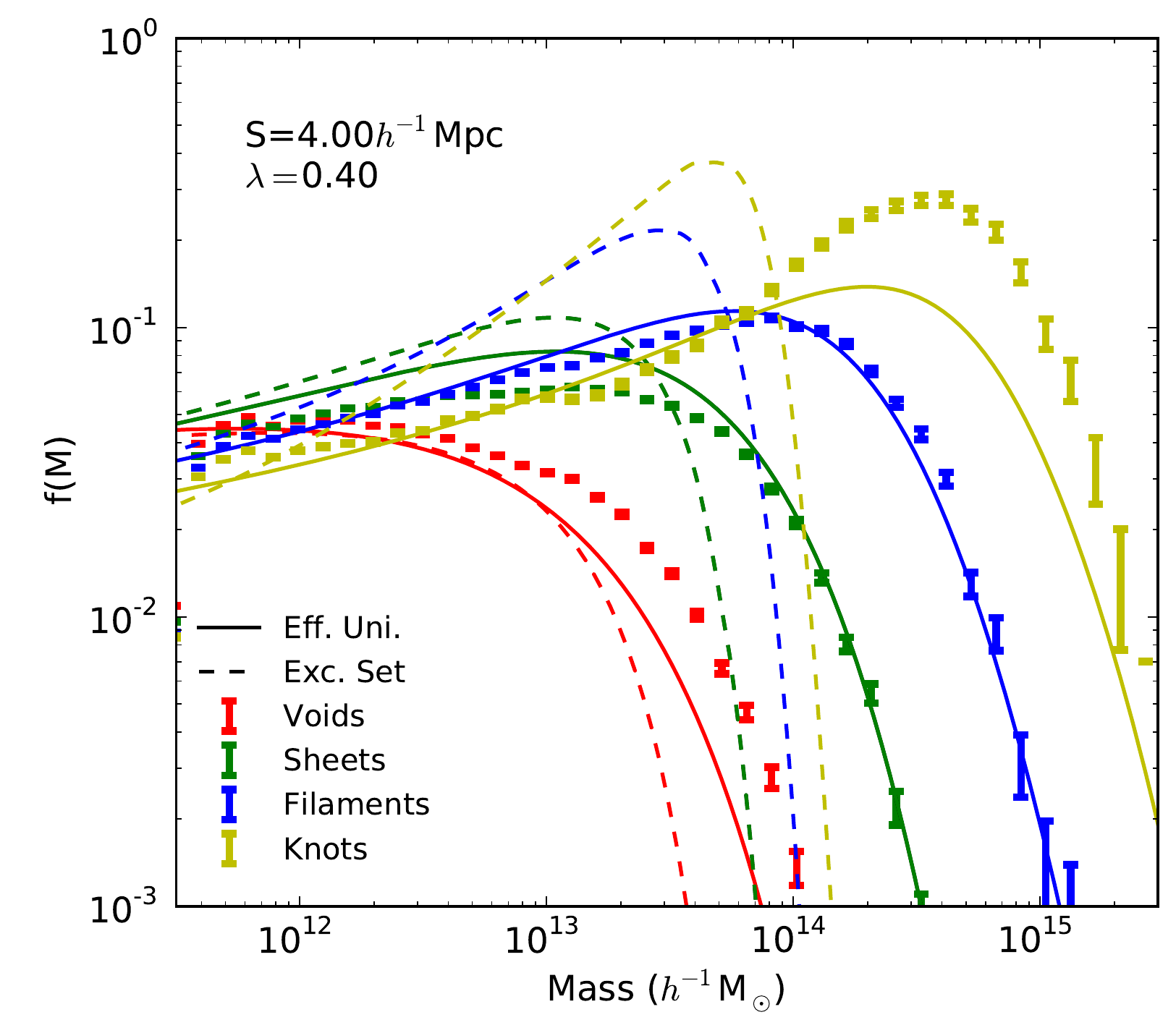}
        \includegraphics[width=0.48\textwidth]
                        {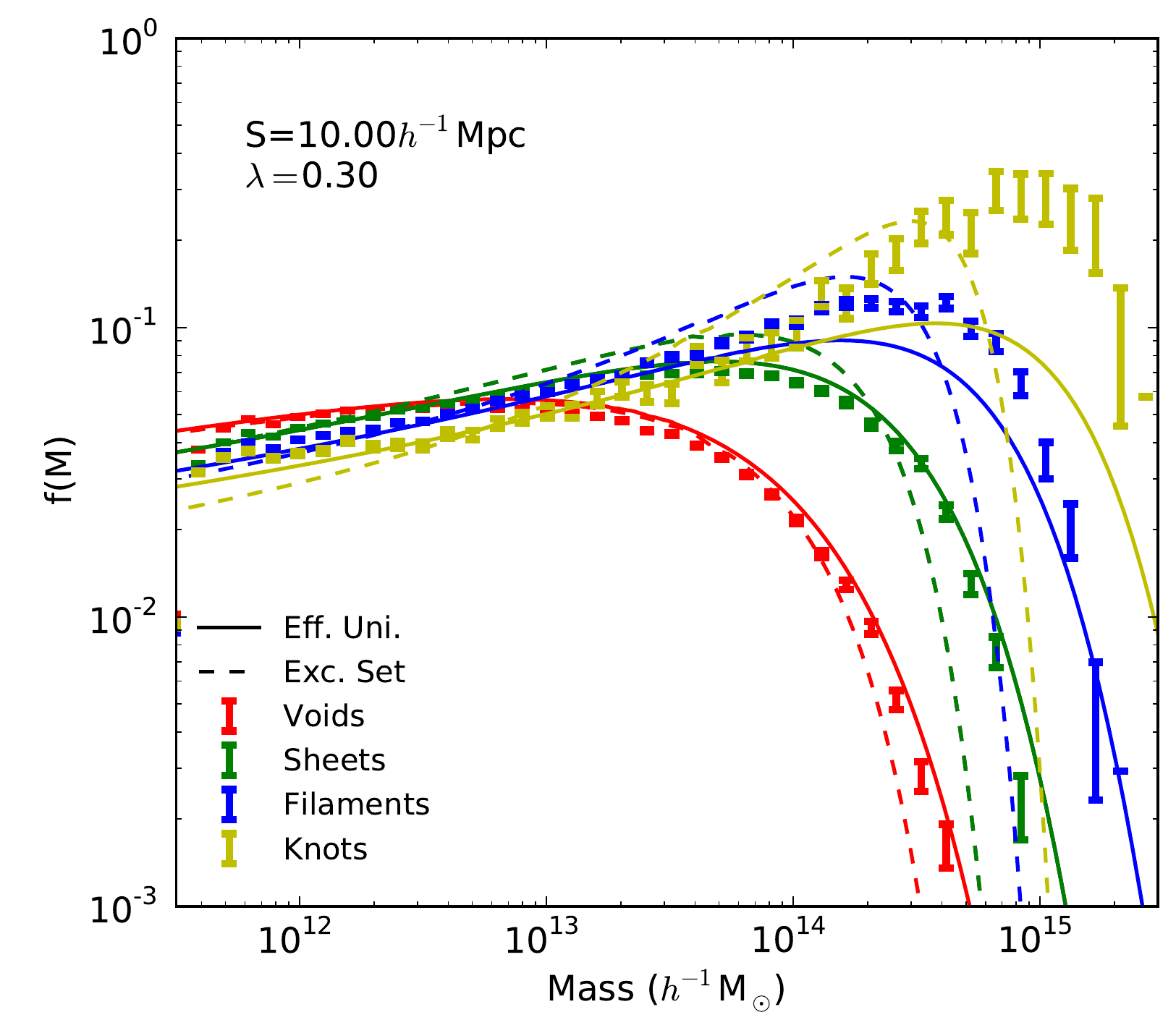}
        \includegraphics[width=0.48\textwidth]
                        {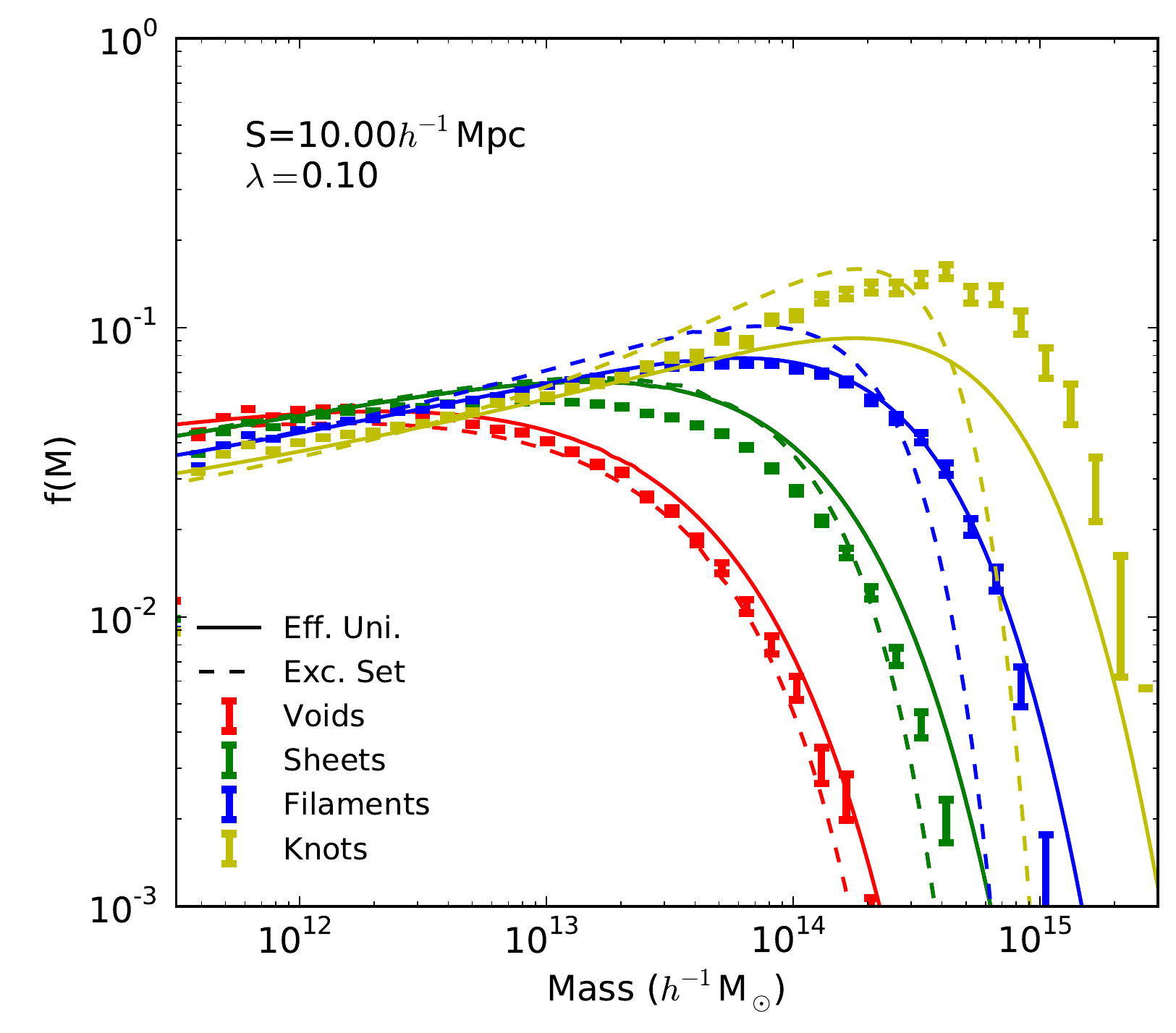}
        \caption{Multiplicity function for the four environments for our fiducial smoothing
                 scales and eigenvalue thresholds. In increasing order of amplitude on high
                 masses, the different multiplicity functions correspond to voids, sheets,
                 filaments and knots. Reasonable agreement is only obtained for large
                 smoothing radii and small $\lambda_{\rm th}$. This can be attributed to the
                 fact that the standard prediction for the conditional mass function is only
                 valid for densities below the collapse threshold and masses below the filter
                 scale. The theoretical prediction was made by rescaling the empirical formula
                 in \citet{2007MNRAS.379.1067P} for the mass function.} \label{fig:mf}
      \end{figure*}
      \begin{figure*}
        \centering
        \includegraphics[width=0.49\textwidth]
          {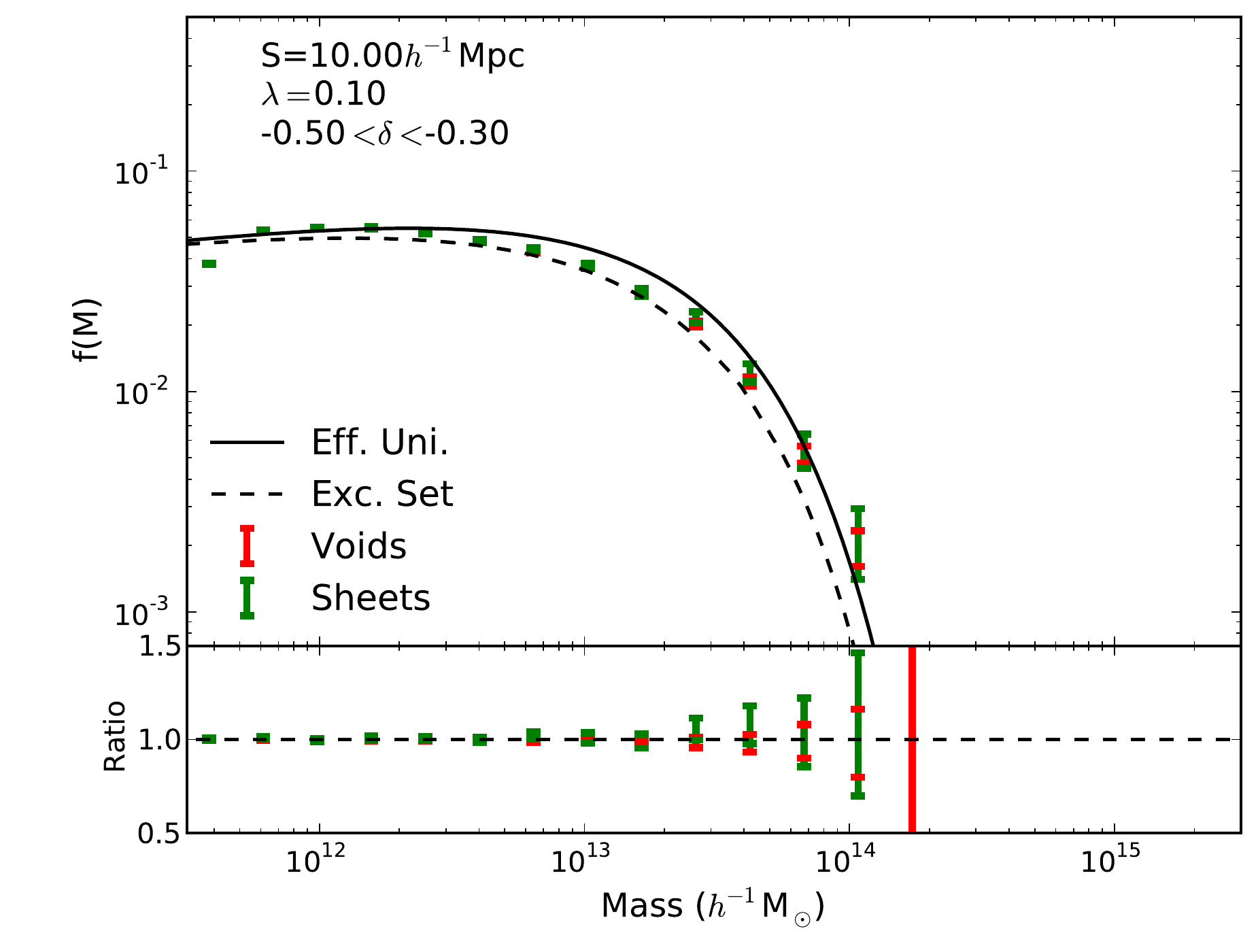}
        \includegraphics[width=0.49\textwidth]
          {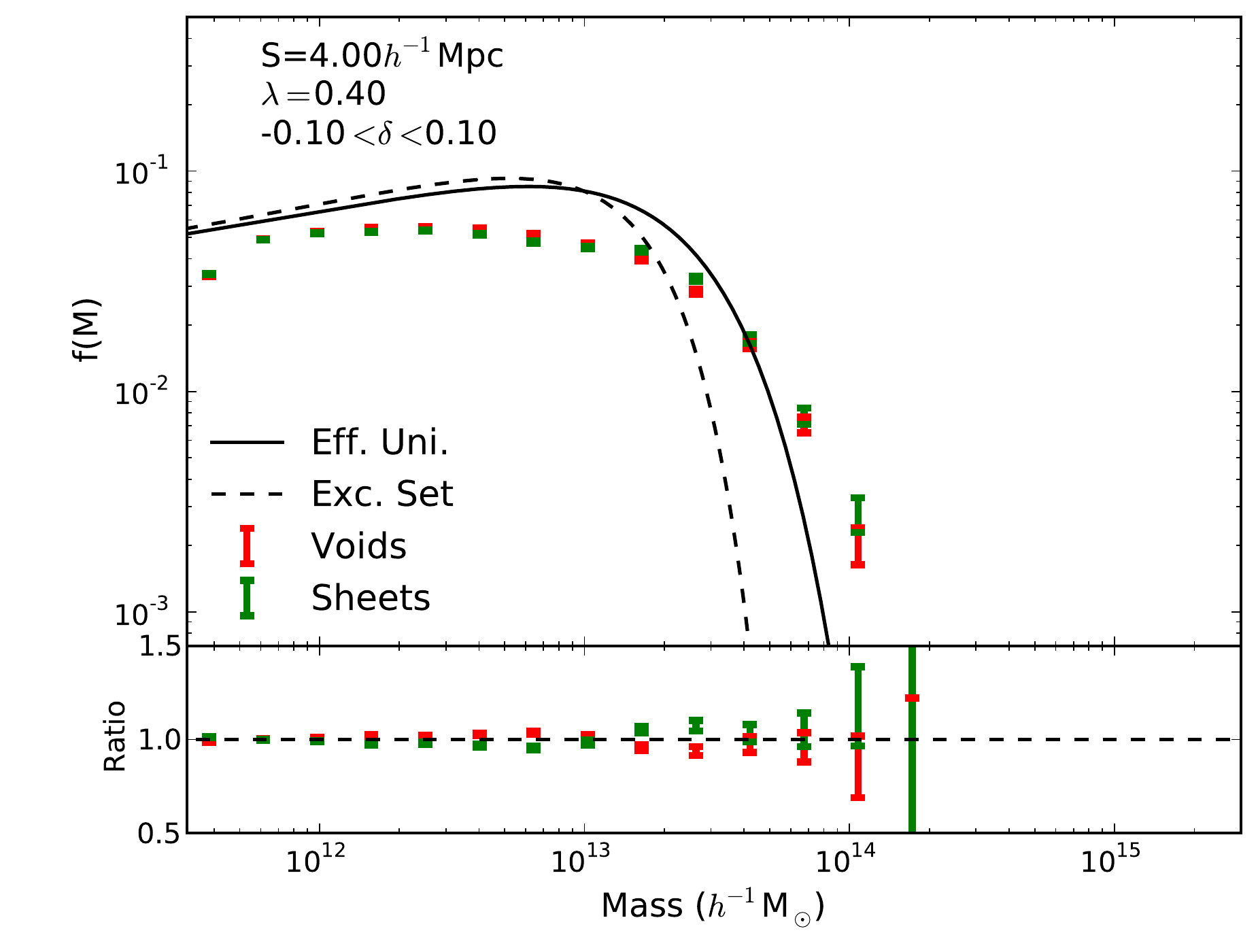}
        \includegraphics[width=0.49\textwidth]
          {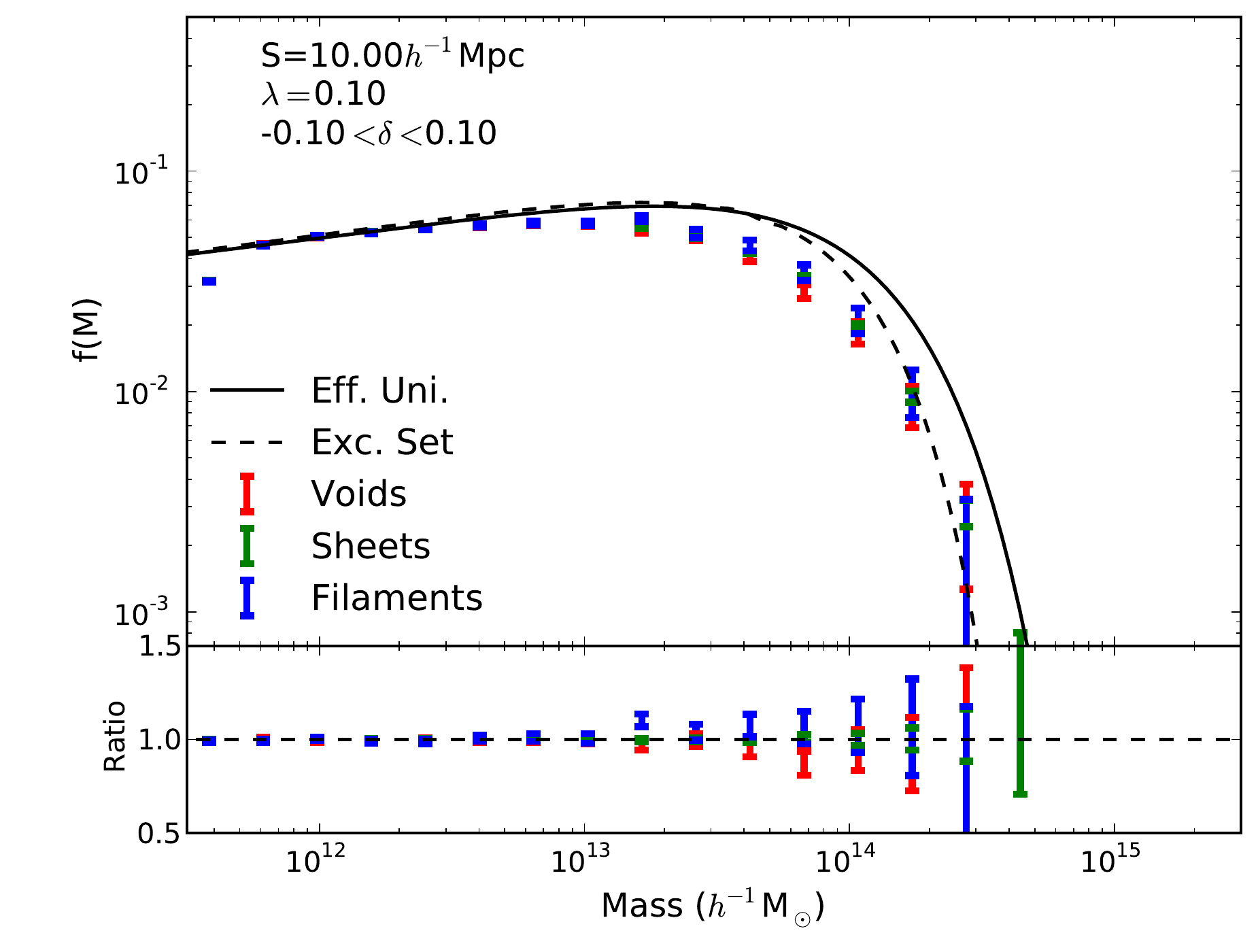}
        \includegraphics[width=0.49\textwidth]
          {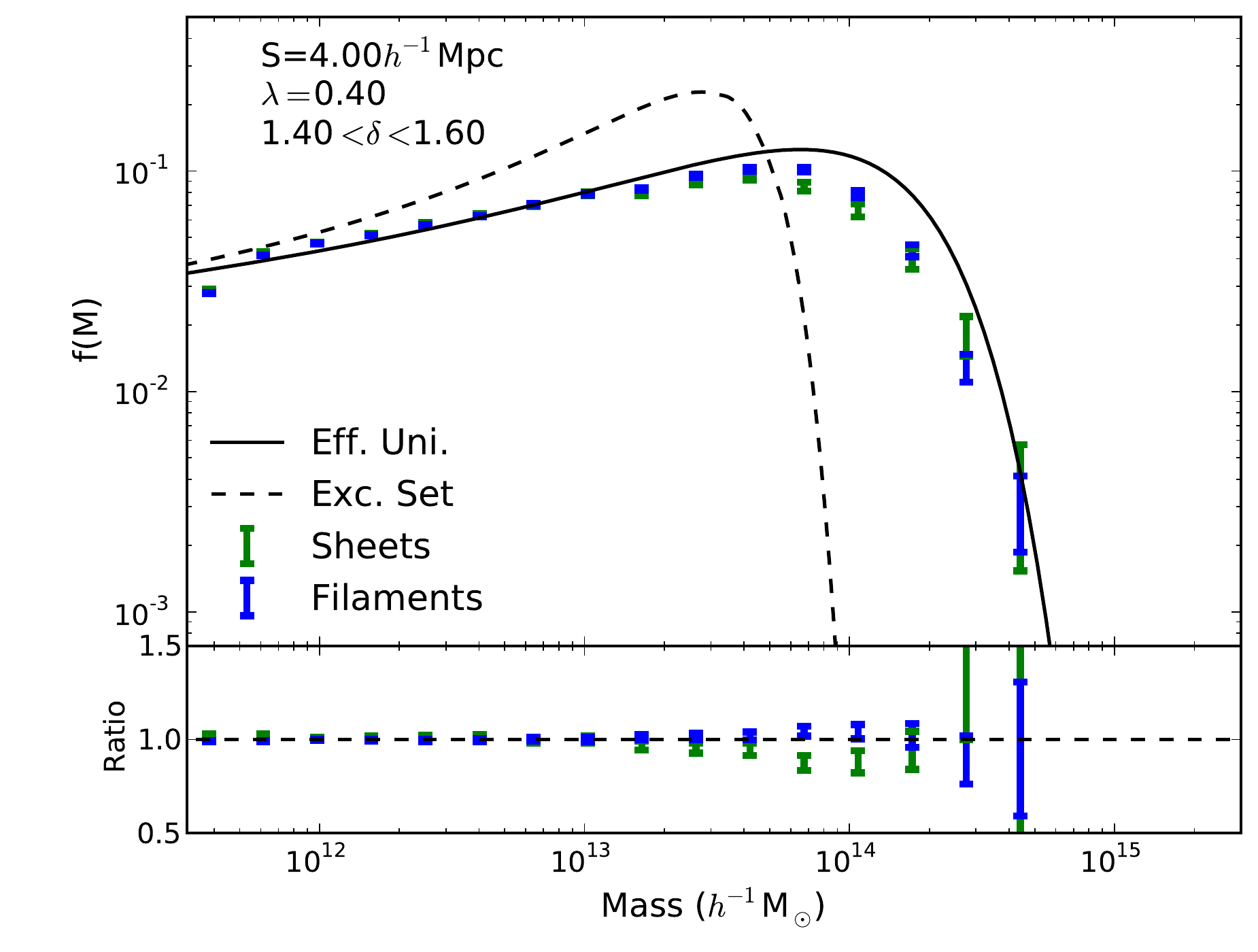}
        \includegraphics[width=0.49\textwidth]
          {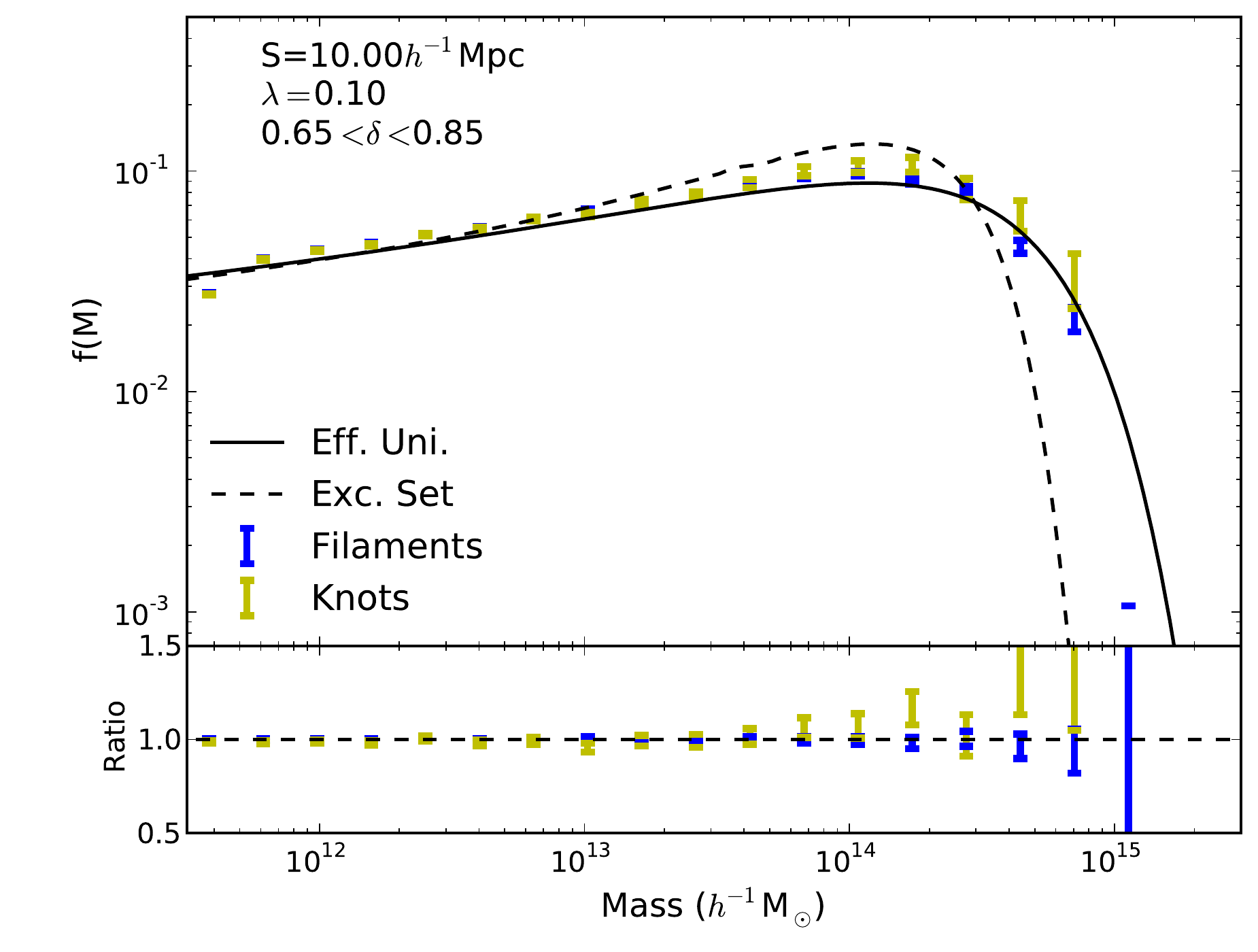}
        \includegraphics[width=0.49\textwidth]
          {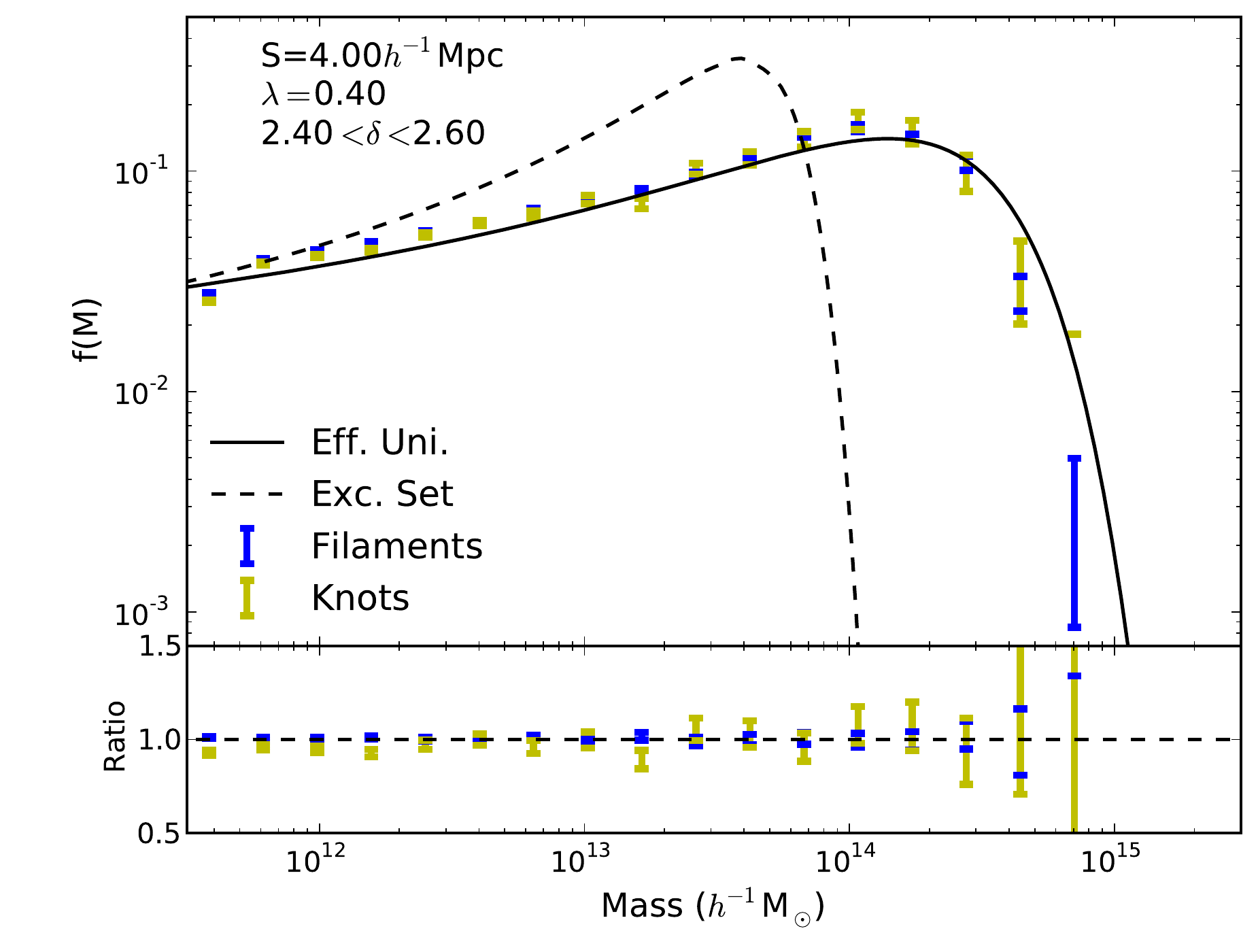}
        \caption{Multiplicity function for the four different environments with their
                 local densities restricted to a given range. In each plot, the lower
                 panel shows the ratio between the different multiplicity functions
                 and their mean value. Our theoretical treatment
                 predicts the same function for all environments in this case, which is
                 realised to a good approximation in all cases. Neither the excursion set
                 prediction nor the effective-universe approach agree quantitatively with
                 the simulated data in all cases, but an overall better agreement is obtained
                 for the effective-universe formalism, especially for smaller scales and larger
                 environmental overdensities.} \label{fig:mf_const_d}
      \end{figure*}

    \subsection{Conditional density distributions}\label{ssec:dens_dist}
      According to the theoretical framework described in Section \ref{sec:stats},
      the halo mass function should be the same in environments with the same density,
      depending only implicitly on the environment classification due to the different
      distribution of densities for each environment type. Therefore, our ability to
      predict the abundance of haloes in each element of the cosmic web depends, on
      the one hand, on reproducing these distributions correctly and, on the other,
      on the accuracy of our model for the conditional mass function.
      These distributions are given, according to our formalism, by
      \begin{equation}
        P(\delta_e|\alpha,\nu_{\rm th})d\delta_e\equiv \frac{P(\nu_e,\alpha,\nu_{\rm th})}
        {F_V(\alpha,\nu_{\rm th})}d\nu_e,
      \end{equation}
      where $P(\nu_e,\alpha,\nu_{\rm th})$ is given by Equation (\ref{eq:cond_pdf}).

      This quantity is shown in Fig. \ref{fig:delta_dist} as measured from the
      MDR1 Simulation for different values of $S$ and $\lambda_{\rm th}$.
      The 2 top panels in this figure show how the density distributions measured
      directly from the simulation (continuous histogram) are very poorly described
      by the Gaussian theory, especially for high-density environments. This is 
      not so surprising, since it is well known that the matter density is
      significantly nonlinear on our filtering scales.
      In order to understand these differences better, we
      have compared these results with the predictions of the lognormal distribution
      \citep{Coles:1991if}, which has often been used as a convenient approximate
      model for the nonlinear density field. Specifically, we 
      used the real density field to perform the environment classification, and
      then we studied the distribution within each environment of the field that
      results from undoing the lognormal transformation. This is given in terms
      of the real overdensity $\delta_r$ and its variance $\sigma_r^2$ as
      \begin{equation}\label{eq:delog}
        \delta_G=\ln\left[(1+\delta_r)\,\sqrt{1+\sigma_r^2}\right].
      \end{equation}

      The distribution of this `de-lognormalized' overdensity is shown in the
      bottom panels of Fig. \ref{fig:delta_dist}. We can see that this field
      follows the Gaussian theoretical distribution much better for the largest
      smoothing scale $S=10\mpcoh$. For smaller scales, however, the
      lognormal transformation is no longer a good description of the non-Gaussianity
      of the density field, a result that has been reported by other
      authors \cite[e.g.][]{2010MNRAS.403..589K}.

    \subsection{Halo abundances} \label{sec:sims_cmf}
      The overall multiplicity function in voids, sheets, filaments and knots is
      calculated by averaging $f(M|\delta_e)$ over $\delta_e$ using the overdensity
      distribution of each environment. As we showed in the previous section, the
      Gaussian prediction for the density distribution $P(\delta_e|\alpha,\nu_{\rm th})$ is
      not a good description of the real distribution in most cases, even after attempting to
      account for non-Gaussianity using the lognormal approximation. Hence, even if we had an
      accurate model for the conditional mass function, we would still not be able to predict
      $f(M|\alpha,\nu_{\rm th})$ correctly. For this reason, in order to isolate the inaccuracies
      due to the incorrect modelling of the conditional mass function from those due to the
      Gaussian approximation, we have integrated over the actual density distributions measured
      from the MDR1 simulation (solid lines in Fig. \ref{fig:delta_dist}) in order to obtain
      a theoretical prediction for the four mass functions. These are shown in Fig. \ref{fig:mf}
      for the cases quoted in Equation (\ref{eq:envirs}) together with the theoretical predictions
      for the excursion set and the effective-universe approaches. These predictions are based on
      rescaling the universal collapsed fraction, which was estimated using the fitting formula
      proposed by \citet{2007MNRAS.379.1067P}. Our results are in qualitative agreement
      with \citet{2007MNRAS.375..489H}, who used an eigenvalue threshold $\lambda_{\rm th}=0$.
      As is discussed in the next section, the excursion set formalism is only able to
      make reasonable predictions for environments involving small overdensities (i.e. voids and
      sheets) and for large filter scales, while the effective-universe approach shows an overall
      better agreement with the data. Nevertheless, neither model is able to describe
      the data accurately. For the present, we have to be content that we have an
      approximate understanding of the trends in halo properties with environment;
      accurate work will require calibration from  numerical simulation, just
      as with the original PS mass function.

    \subsection{Universality of density dependence}
      The above results show that the excursion set approach is able to make relatively good
      predictions for large smoothing scales and mild environmental densities, but that it fails to
      do so for smaller values of $S$ and high $\delta_e$. This is a reasonable result:
      the excursion set model is based on following trajectories in the $\delta - R$
      plane that, starting at some $(R_e,\delta_e)$, cross the threshold $\delta_c$ at
      some scale $R(M)<R_e$. In this regime the excursion set predicts that too many
      small-mass haloes have already merged into larger ones, due to the fact that the
      large environmental density makes gravitational collapse more efficient. At the
      large-mass end, on the other hand, too few haloes have formed, since the total halo
      mass is limited by the mass that can be found within the smoothing radius. Also,
      for small $R_e$ the correlation between adjacent steps,
      which is generally ignored, may play a significant role, since the scale of the
      halo mass can be close to the scale of the environment. In these limits, the
      effective-universe approach outperforms the excursion set, providing a better
      description for the conditional mass function -- although far from per-cent level
      precise.

      Nevertheless, one of the most important predictions of our formalism is that the abundance
      of haloes should be the same for all environments with the same background density.
      Regardless of whether or not the theoretical prediction for the conditional mass function
      is quantitatively precise, it is interesting to test the qualitative validity
      of this result according to the simulated data.

      Fig. \ref{fig:mf_const_d} shows the multiplicity function ($f(M)\equiv\left|d\,F(<M)/d\,
      \ln{M}\right|$) for haloes residing in the four different environments with restricted local
      overdensities, together with the environment-independent theoretical predictions derived
      from the excursion set (dashed lines) and the effective-universe approach
      (solid lines) as detailed in Section \ref{sec:cmf}. The values of
      the local overdensity were chosen to guarantee the simultaneous presence of
      as many different environments as possible. For the range of masses, smoothing scales and
      densities explored, we find that the prediction that the abundance of haloes should depend
      only on the environmental density and not on the environment classification holds very well,
      with little or no deviation within statistical errors. We have quantified this agreement
      as follows: for each overdensity bin we take all pairs of multiplicity
      functions that we have been able to calculate in it. Assuming Gaussian statistics, we
      estimate the probability that both multiplicity functions are compatible in each mass bin
      given their statistical uncertainties. We then quantify the agreement between multiplicity
      functions by computing the relative difference between them in the
      mass bin with the smallest $p$-value. The results are shown in Table \ref{tab:comp} for the
      bins of overdensity explored above. In the worst case, the largest deviation is about 13\%.
      However it is worth noting that, even in this case, both multiplicity functions are fully
      compatible, with a minimum $p$-value of $0.32$, and in all other cases the values of the
      different multiplicity functions are compatible within 2-$\sigma$.
      
      Even though we have verified the prediction that the halo abundances in different
      environments depend only on the environmental density, the exact dependence of these
      abundances on halo mass is not reproduced accurately by either the effective-universe
      approach or the excursion set formalism, although they both qualitatively follow the same
      trend as the data. 
      \begin{table}
        \begin{center}
          \begin{tabular}{c|c||c|c}
              \multicolumn{2}{c||}{$S=10\,h^{-1}{\rm Mpc},\,\lambda_{\rm th}=0.1$} &
              \multicolumn{2}{c}{$S=4\,h^{-1}{\rm Mpc},\,\lambda_{\rm th}=0.4$}\\
              \hline
              $\delta$ & $\Delta f(M)\,(\%)$ & $\delta$ & $\Delta f(M)\,(\%)$\\
              \hline
              $(-0.5,-0.3)$ & 1.1\% & $(-0.1,0.1)$ & 4.6\% \\
              $(-0.1,0.1)$ & 13.2\% & $(1.4,1.6)$ & 3.5\% \\
              $(0.65,0.85)$ & 5.2\% & $(2.4,2.6)$ & 8.6\% \\
              \hline
            \end{tabular}
          \end{center}
          \caption{Compatibility of the mass functions for different environments with
                   restricted environmental densities for the two combinations of
                   $(S,\lambda_{\rm th})$ explored in Figure  \ref{fig:mf_const_d}:
                   $(10\,h^{-1}{\rm Mpc},0.1)$, left column, and
                   $(4\,h^{-1}{\rm Mpc},0.4)$, right column.}
          \label{tab:comp}
       \end{table}
       
  \section{Summary and Discussion}\label{sec:discussion}
    We have considered the statistics of dark-matter haloes within
    the cosmic web, using the eigenvalues of the Hessian of the
    potential to classify regions of space into one of four
    geometrical environments. The main results from this work are:
    \begin{enumerate}
      \item Assuming the density contrast field to be Gaussian, clear
            predictions can be made regarding the abundance (i.e. volume and mass
            fractions) of the different environments classified according to the
            tidal tensor prescription. These are reasonable approximations for 
            large smoothing scales and can be used to select
            eigenvalue thresholds that are useful for practical comparisons -- 
            partitioning the Universe nearly equally between the four environments.
      \item We have compared the simulated halo abundances with the predictions within
            the excursion set formalism and the effective-universe approach. Neither of these
            approaches are able to yield quantitatively precise results. However, the
            effective-universe picture provides an overall better description, especially for
            small smoothing scales and large environmental densities.
      \item The Gaussian approach predicts that the only local property of the environment
            on which the conditional mass function depends is the density contrast
            $\delta_e$. Thus a prediction of the mass function in the different geometrical
            environments should be possible if we know the overall dependence of the mass
            function on overdensity, and if we can predict the overdensity distributions for
            the different environments.
      \item A detailed test of this prediction does not succeed very well, since the overdensity
            distributions are not well predicted by the Gaussian theory. This is improved in part
            by considering a lognormal model for the evolved density field, but discrepancies
            with numerical data remain.
      \item Nevertheless, we have been able to test directly the fundamental prediction of this
            work, which is that different geometrical environments with the same overdensity
            should have identical halo mass functions. This is verified in the MDR1 simulation to
            a good approximation for a wide range of masses, filters and eigenvalue thresholds.
            We find a maximum relative deviation of aboutt 13\% between mass functions in different
            environments, which are, nevertheless, fully compatible in a statistical sense.
            In this regard, we see no evidence for any effect of tidal forces on halo abundances
            in addition to the impact of local overdensity. This could be consistent with the claim
            by \citet{Yan2013} that galaxy properties in the SDSS lacked an explicit dependence on
            envionmental ellipticity, as well as with the results found by
            \citet{2015MNRAS.446.1458M} in N-body simulations; it will be interesting to repeat
            such an analysis using the explicit decomposition by geometrical environment that we
            have studied here.
      \item This result suggests that scalar halo properties are not heavily influenced by the
            tidal field beyond the local overdensity. This is not at odds with the results found
            by \citet{2014arXiv1406.0508F,2014arXiv1407.0394L}, who find that directional
            quantities, such as orientations or angular momenta, show a strong correlation with
            the directions defined by the tidal tensor.
    \end{enumerate}

  \section*{Acknowledgements}
    We would like to thank Catherine Heymans and Juan Garc\'ia-Bellido for useful comments and
    discussions. DA is supported by ERC grant 259505 and acknowledges the hospitality of the Royal
    Observatory, Edinburgh, during a research stay supported by a JAE-Predoc scholarship. EE is
    supported by a STFC postgraduate studentship. The MultiDark Database used in this paper and
    the web application providing online access to it were constructed as part of the activities
    of the German Astrophysical Virtual Observatory as result of a collaboration between the
    Leibniz-Institute for Astrophysics Potsdam (AIP) and the Spanish MultiDark Consolider Project
    CSD2009-00064. The Bolshoi and MultiDark simulations were run on the NASA's Pleiades
    supercomputer at the NASA Ames Research Center. The MultiDark-Planck (MDPL) and the BigMD
    simulation suite have been performed in the Supermuc supercomputer at LRZ using time
    granted by PRACE.

  \setlength{\bibhang}{2.0em}
  \setlength{\labelwidth}{0.0em}
  \bibliography{paper}

  \appendix
  \section[]{The effective universe approach}\label{app:eff_univ}
    Consider a spherical perturbation in an otherwise homogeneous universe. It is a
    well-known result in gravitational theory that at any distance from the centre
    of the perturbation, it must evolve as a parallel FRW cosmology with some effective
    cosmological parameters which can be entirely determined in terms of the 
    amplitude of the density perturbation. This result allows us to interpret the
    conditional mass function for an environment with overdensity $\delta_e$ as the
    mass function in the corresponding effective universe.

    The `environmental' cosmological parameters are related to the background ones
    and the perturbation's overdensity through:
    \begin{equation}
      \Omega_m^e=\Omega_m^{\rm BG}\,(1+\delta)/\eta^2\,\,\,\,
      \Omega_{\Lambda}^e=\Omega_{\Lambda}^{\rm BG}/\eta^2\,\,\,\,
      H_0^e=\eta^2\,H_0^{\rm BG},
    \end{equation}
    where the superscripts ${\rm BG}$ and $e$ denote quantities in the background and
    in the effective universe respectively. The ratio between the current expansion
    rates inside and outside the perturbation, $\eta$, can be fixed by imposing that
    the age of the Universe
    \begin{equation}
      t_{\rm BB}=\frac{1}{H_0}\int_0^1\frac{dx}{x\sqrt{\Omega_mx^{-3}+
                                            \Omega_{\Lambda}+\Omega_kx^{-2}}}
    \end{equation}
    must be the same as measured by any observer. This effectively implies that the
    perturbation must be a purely growing mode that disappears at early times.

    Once the effective cosmological parameters are known, the scaling factor
    $D_g$ in Equation (\ref{eq:eff_var}) is given by the ratio of the growth
    factors in the two cosmologies. Normalising this ratio to be 1 at early
    times (where the perturbation gradually disappears), this quantity is
    given by
    \begin{equation}
      D_g=\left(\frac{\Omega_m^{\rm BG}h_{\rm BG}^2}{\Omega_m^eh_e^2}\right)^{1/3}
      \frac{\Omega_m^e}{\Omega_m^{\rm BG}}\frac{g(\Omega_m^e,\Omega_{\Lambda}^e)}
      {g(\Omega_m^{\rm BG},\Omega_{\Lambda}^{\rm BG})},
    \end{equation}
    where
    \begin{equation}
      g(\Omega_m,\Omega_{\Lambda})=\int_0^1dx
      \left(\frac{x}{\Omega_m+\Omega_{\Lambda}x^3+\Omega_kx}\right)^{3/2}.
    \end{equation}

    Notice that at this point we have not taken into account the size of the environment.
    For large smoothing scales or comparatively small halo masses, this is not an
    important concern: we may treat the environment as an infinite effective universe in
    which haloes of any mass may form. In practice however, the mass of the largest haloes
    ($M\sim10^{15}\,M_{\odot}\rightarrow R_h\sim15 {\rm Mpc}$) corresponds to
    scales of the order of the filter scale used to define the environment,
    and hence halo masses must be restricted by the amount of matter that
    is available in their environment. We have taken this effect into
    account by restricting the Fourier modes that can contribute to the
    variance of the overdensity field in a given environment, suppressing 
    those corresponding to scales larger than $R_e$. In practice we have
    implemented this by weighting each mode by the `inverse' of the
    window function used to define the environment, $W_e$:
    \begin{equation}\label{eq:var_corrected}
      \sigma_{\rm eff}(M)=\int_0^{\infty}\frac{k^2dk}{2\pi^2}\,\left[1-W_e(kR_e)\right]^2
                      |W(kR_h)|^2P_k.
    \end{equation}

  \section[]{Correlations between environment and density for Gaussian fields}\label{app:cweb}
    \subsection{The eigenvalue distribution}
      Consider a Gaussian potential field $\tilde{\phi}$ smoothed over a
      length scale $R$. Since the tidal tensor $\hat{T}$ is a symmetric matrix only 6 of its
      components are independent. We will label them with a single index: $T_A=(T_{11},
      T_{22},T_{33},T_{23},T_{31},T_{12})$. It is straightforward to calculate the covariance
      matrix of the $T_A\,'s$:
      \begin{equation}
        \langle T_AT_B\rangle=\frac{\sigma_R^2}{15}\left(
			\begin{array}{cccccc}
			 3 & 1 & 1 & 0 & 0 & 0 \\
			 1 & 3 & 1 & 0 & 0 & 0 \\
			 1 & 1 & 3 & 0 & 0 & 0 \\
			 0 & 0 & 0 & 1 & 0 & 0 \\
			 0 & 0 & 0 & 0 & 1 & 0 \\
			 0 & 0 & 0 & 0 & 0 & 1
			\end{array}\right),
      \end{equation}
      where $\sigma_R^2$ is given in Equation (\ref{eq:sigmavar}) with $R_a=R_b=R$. This matrix
      can be diagonalized by changing to the variables:
      \begin{align}
        \nonumber
        & \tau_1\equiv\nu\equiv\frac{1}{\sigma_R}(T_1+T_2+T_3),\,\,\,
          \tau_2\equiv\rho\equiv\frac{1}{2\sigma_R}(T_1-T_3),\\
          \nonumber
        & \tau_3\equiv\theta\equiv\frac{1}{2\sigma_R}(T_1-2T_2+T_3),\\
        & \tau_4\equiv\frac{T_4}{\sigma_R},\,\,\,
          \tau_5\equiv\frac{T_5}{\sigma_R},\,\,\,
          \tau_6\equiv\frac{T_6}{\sigma_R}.
      \end{align}
      Notice that by definition, $\nu$ is proportional to the local density contrast:
      $\delta=\nu\sigma_R$ and that $\rho$ and $\theta$ are trivially related to
      the ellipticity $e\equiv\rho/\nu$ and prolateness $p\equiv\theta/\nu$. In
      terms of these new variables the covariance matrix is diagonal:
      \begin{equation}
        \mu_{AB}\equiv\langle \tau_A\tau_B\rangle=
        {\rm diag}\left(1,\frac{1}{15},\frac{1}{5},\frac{1}{15},\frac{1}{15},
        \frac{1}{15}\right)
      \end{equation}
      and the joint distribution of the $T_A$'s is
      \begin{align}
        \nonumber
        P(\{T_A\})\prod_AdT_A=\frac{e^{-Q/2}}{\sqrt{(2\pi)^6\,{\rm det}(\hat{\mu})}}
        \prod_Ad\tau_A,\\
        Q\equiv \nu^2+15\rho^2+5\theta^2+15(\tau_4^2+\tau_5^2+\tau_6^2).
      \end{align}
      This holds in any coordinate system, but it will be most useful in the one
      in which $\hat{T}$ is diagonal (i.e. $\hat{T}=
      \,{\rm diag}(\lambda_1,\, \lambda_2,\,\lambda_3,\,0,0,0)$). As proved
      by \citet{1986ApJ...304...15B} the volume element of the space of $3\times3$
      symmetric matrices can be written in terms of the matrix eigenvalues
      and the Euler angles of the rotation necessary to diagonalise it:
      \begin{equation}
        \prod_AdT_A=|(\lambda_1-\lambda_2)(\lambda_2-\lambda_3)(\lambda_1-\lambda_3)|\,
        d\lambda_1\,d\lambda_2\,d\lambda_3
        \frac{d\Omega_3}{6},
      \end{equation}
      where $d\Omega_3$ is the volume element of $\mathcal{S}^3$ (the total volume of
      which is $2\pi^2$). Up to now we have not chosen any specific ordering
      for the eigenvalues.  There are 6 possible orderings, and the probability
      density is symmetric with respect to these, therefore imposing a specific
      ordering would introduce a factor $6$ in the probability density above.
      Choosing $\lambda_1\ge\lambda_2\ge\lambda_3$ (which implies
      $\theta\in[-\rho,\rho],\,\,\,\rho\in[0,\infty)$) and integrating out
      the irrelevant angular part we obtain the probability distribution given
      in Equation (\ref{eq:pdf_lambdas}). 

    \subsection{Correlation with the local density contrast}
      Now let us consider the distribution of the density contrast smoothed
      over a scale $R_h$ in different regions classified via the tidal
      tensor eigenvalues smoothed over a scale $R_e$. Here we will use the
      notation introduced in Section \ref{sec:cmf}, with
      $\delta_e\equiv{\rm Tr}(\hat{T})$.

      The covariance of $\delta_h$ with the $T_A$'s is easy to calculate:
      \begin{equation}
        \langle \delta_h\,T_A\rangle=\frac{\sigma_{eh}^2}{3}(\delta_{A1}+\delta_{A2}+
        \delta_{A3})
      \end{equation}
      Defining $G_a=(\delta_h,T_A),\,\,\,\,(a=0,...,6)$ the full covariance matrix is
      \begin{align}
        \nonumber
        &\langle G_aG_b\rangle=\frac{1}{15}\times\\
        &\left(\begin{array}{ccccccc}
          15\sigma_{hh}^2 & 5\sigma_{eh}^2 & 5\sigma_{eh}^2 & 5\sigma_{eh}^2 & 0 & 0 & 0 \\
          5\sigma_{eh}^2 & 3\sigma_{ee}^2 & \sigma_{ee}^2 & \sigma_{ee}^2 & 0 & 0 & 0 \\
          5\sigma_{eh}^2 & \sigma_{ee}^2 & 3\sigma_{ee}^2 & \sigma_{ee}^2 & 0 & 0 & 0 \\
          5\sigma_{eh}^2 & \sigma_{ee}^2 & \sigma_{ee}^2 & 3\sigma_{ee}^2 & 0 & 0 & 0 \\
          0 & 0 & 0 & 0 & \sigma_{ee}^2 & 0 & 0 \\
          0 & 0 & 0 & 0 & 0 & \sigma_{ee}^2 & 0 \\
          0 & 0 & 0 & 0 & 0 & 0 & \sigma_{ee}^2
        \end{array}\right).
      \end{align}
      As before, in order to simplify this matrix we define:
      \begin{align}
        \nonumber
        & \gamma_0\equiv\nu_h\equiv\frac{\delta_h}{\sigma_{hh}},\,\,\,
          \gamma_1\equiv\nu_e\equiv\frac{T_1+T_2+T_3}{\sigma_{ee}},\\
          \nonumber
        & \gamma_2\equiv\rho\equiv\frac{T_1-T_3}{2\sigma_{ee}},\,\,\, 
          \gamma_3\equiv\theta\equiv\frac{T_1-2T_2+T_3}{2\sigma_{ee}},\\
        & \gamma_4\equiv\frac{T_4}{\sigma_{ee}},\,\,\,
          \gamma_5\equiv\frac{T_5}{\sigma_{ee}},\,\,\,
          \gamma_6\equiv\frac{T_6}{\sigma_{ee}},
      \end{align}
      the covariance of which is
      \begin{equation}
        \langle \gamma_a\gamma_b\rangle=\left(
			\begin{array}{ccccccc}
			 1 & \varepsilon & 0 & 0 & 0 & 0 & 0 \\
			 \varepsilon & 1 & 0 & 0 & 0 & 0 & 0 \\
			 0 & 0 & 1/15 & 0 & 0 & 0 & 0 \\
			 0 & 0 & 0 & 1/5 & 0 & 0 & 0 \\
			 0 & 0 & 0 & 0 & 1/15 & 0 & 0 \\
			 0 & 0 & 0 & 0 & 0 & 1/15 & 0 \\
			 0 & 0 & 0 & 0 & 0 & 0 & 1/15
			\end{array}\right),
      \end{equation}

      Now we can follow the same procedure as before: change the volume element
      of the $T_A$'s to the one in the space of eigenvalues and rotations,
      change to the coordinate system in which $\hat{T}$ is diagonal,
      transform everything to our variables $\{\gamma_a\}$, choose a
      specific ordering for the eigenvalues and integrate out the irrelevant
      angular part. At the end of the day we obtain the distribution
      given in Equation (\ref{eq:cond_pdf}).

\end{document}